\documentclass{aa}  

\newcounter{daggerfootnote}

\usepackage{graphicx}

\usepackage{txfonts}
\usepackage{natbib}
\usepackage[switch]{lineno}

\usepackage{xspace}
\usepackage{color}

\newcommand{\kms}     {km~s$^{-1}$\xspace}
\newcommand{\au}      {$\mathrm{au}$}
\newcommand{\yr}      {$\mathrm{yr^{-1}}$}
\newcommand{\msun}    {$M_{\sun}$}

\newcommand{\mjy}     {~mJy~beam$^{-1}$\xspace}

\begin{document} 

\setlength{\linenumbersep}{10pt}

   \titlerunning{} 
   \authorrunning{L\'opez-V\'azquez, J. A.}
   \title{Erosion of a dense molecular core by a strong outflow from a massive protostar}


   \author{J. A. L\'opez-V\'azquez
          \inst{1},\, 
          M. Fern\'andez-L\'opez\inst{2},\, 
          J. M. Girart\inst{3,4},\, S. Curiel\inst{5},\, R. Estalella\inst{6,\dagger},\,G. Busquet\inst{6,7},\, Luis A. Zapata\inst{8},\,Chin-Fei Lee\inst{1},\, \and Roberto Galv\'an-Madrid\inst{8}
          }


   \institute{Academia Sinica Institute of Astronomy and Astrophysics, No. 1, Sec. 4, Roosevelt Road, Taipei 10617, Taiwan
   \and Instituto Argentino de Radioastronom\'ia (CCT-La Plata, CONICET; UNLP; CICPBA), C.C. No. 5, 1894, Villa Elisa, Buenos Aires, Argentina
   \and Institut de Ciencies de l’Espai (ICE-CSIC), Campus UAB, Carrer de Can Magrans S/N, E-08193 Cerdanyola del Valles, Catalonia
   \and Institut d'Estudis Espacials de Catalunya (IEEC), c/ Gran Capità, 2-4, 08034 Barcelona, Spain
   \and Instituto de Astronom\'{i}a, Universidad Nacional Aut\'onoma de M\'exico (UNAM), Apartado Postal 70-264, DF 04510 M\'exico
   \and Departament de Física Quàntica i Astrofísica (FQA), Universitat de Barcelona (UB), c/ Martí i Franquès 1, 08028 Barcelona, Spain
   \and Institut de Ci\'encies del Cosmos (ICCUB), Universitat de Barcelona (UB), c/ Martí i Franquès 1, 08028 Barcelona, Spain
   \and
   Instituto de Radioastronom\'ia y Astrof\'isica, Universidad Nacional Aut\'onoma de M\'exico, Apartado Postal 3-72, 58089 Morelia, Michoac\'an, M\'exico
             }


 
  \abstract
   {Molecular outflows from massive protostars can impact the Interstellar Medium in different ways, adding turbulence in different spatial scales, dragging material at supersonic velocities, producing shocks and heating, and also physically impinging onto dense structures that may be harbouring other protostars.}
   {We aim to quantify the impact of the outflow associated with the high-mass protostar GGD~27-MM2(E) on its parent envelope and how this outflow affects its environment.}
   {We present Atacama Large Millimeter/submillimeter Array Band 3 observations of N$_2$H$^+$~(1-0) and CH$_3$CN~(5-4), as well as Band 7 observations of the H$_2$CO molecular line emissions from the protostellar system GGD~27-MM2(E). Through position-velocity diagrams along and across the outflow axis, we study the kinematics and structure of the outflow. We also fit extracted spectra of the CH$_3$CN emission to obtain the physical conditions of the gas. We use the results to discuss the impact of the outflow on its surroundings.}
   {We find that N$_2$H$^+$ emission traces a dense molecular cloud surrounding GGD~27-MM2(E). We estimate that the mass of this cloud is $\sim$13.3--26.5 \msun. The molecular cloud contains an internal cavity aligned with the H$_2$CO-traced molecular outflow. The outflow, also traced by $\mathrm{CH_3 CN}$, shows evidence of a collision with a molecular core (MC), as indicated by the distinctive increases in the distinct physical properties of the gas such as excitation temperature, column density, line width, and velocity. This collision results in an X-shape structure in the northern part of the outflow around the position of the MC, which produces spray-shocked material downstream in the north of MC as observed in position-velocity diagrams both along and across of the outflow axis. The outflow has a mass of 1.7--2.1 \msun, a momentum of 7.8--10.1 \msun\,\kms, a kinetic energy of 5.0--6.6$\times 10^{44}$ erg, and a mass loss rate of 4.9--6.0$\times10^{-4}$ \msun\,\yr. }
   {The molecular outflow from GGD~27-MM2(E) significantly perturbs and erodes its parent cloud, compressing the gas of sources such as MC and ALMA~12. The feedback from this powerful protostellar outflow contributes to maintain the turbulence in the surrounding area.}


   \maketitle
%

\section{Introduction}
\label{sec:introduction}
\renewcommand{\thefootnote}{\dag}
\footnotetext{Deceased on November 19, 2024.}
\renewcommand{\thefootnote}{\arabic{footnote}}
Protostellar jets and bipolar outflows typically emerge during the early stages of star formation, when a cold, dusty envelope starts to gather material into a disk.
High accretion rates during this phase quickly lead to the formation of a central core, which will evolve into a protostar \citep[e.g., ][]{Blandford1982,Pudritz1983,Bally2016,Maureira2017}. The protostellar jets represent the fast and collimated wind launched from the inner region of the disk, while the bipolar outflows are the material dragged by the jet or by the outflowing stellar wind \citep[e.g.,][]{Snell1980}. They have previously classified as either a wide-angle disk wind or a jet-driven wind \citep[see e.g.,][]{Lee2000,Lee2001}.

Despite the traditional theoretical expectation that outflows are bipolar, well-behaved structures --backed by a plethora of observations-- there are several cases that do not follow this rule. There are detections of monopolar outflows \citep[e.g.,][]{FernandezLopez2013,Stanke2022,Takaishi2024}, outflows with multiple collimated ejections in different directions \citep[e.g.,][]{Cunningham2009,Soker2013,Kwon2015,Takaishi2024,Sai2024}, precessing jets and outflows \citep{Marti1993,Anglada2007,Teixeira2008,Arce2013,Zapata2013}, and, lately, even detections of their spiraling streamlines using maser emission \citep{Moscadelli2022,Moscadelli2023}. 

\begin{table*}[t!]
    \centering
    \caption{Spectral lines}
    \begin{tabular}{c c c c c c c}
    \hline
    \hline
    Transition & $\nu_\mathrm{rest}$ & $E_{up}$ & \multicolumn{2}{c}{Synthesized Beam}  & $\delta v$ & rms \\
    \cline{4-5}
         & (GHz) & (K) & (arcsec$\times$arcsec) & (deg) & (\kms) & (\mjy) \\
    \hline
    CH$_3$CN (5-4) K=3 & 91.971153 & 77.6 & $1^{\prime\prime}.3\times 1^{\prime\prime}.2$ & 71 & 0.8 & 1.2 \\
    CH$_3$CN (5-4) K=2 & 91.980087 & 41.8 & $1^{\prime\prime}.3\times 1^{\prime\prime}.2$ & 71 & 0.8 & 1.2 \\
    CH$_3$CN (5-4) K=1 & 91.985304 & 20.4 & $1^{\prime\prime}.3\times 1^{\prime\prime}.2$ & 71 & 0.8 & 1.2 \\
    CH$_3$CN (5-4) K=0 & 91.987088 & 13.2 & $1^{\prime\prime}.3\times 1^{\prime\prime}.2$ & 71 & 0.8 & 1.2 \\
    \hline
    N$_2$H$^{+}$ (1-0) & 93.173398 &  4.5 & $1^{\prime\prime}.3\times 1^{\prime\prime}.2$ & 71 & 0.8 & 1.2 \\
    \hline
    H$_2$CO ($4_{1,3}-3_{1,2}$) & 300.836635 & 47.9 & $0.^{\prime\prime}18\times0.^{\prime\prime}13$ & 80 & 0.5 & 2.1 \\
    \hline
    \end{tabular}
    \tablefoot{The quoted rest frequencies have been extracted from the Cologne Database for Molecular Spectroscopy \citep[CDMS,][]{2005Muller}, using the online port Splatalogue (\citealt{Remijan2007}) at \url{https://splatalogue.online/\#/home}. The rest frequency for the N$_2$H$^+$~(1-0) multiplet is the average (weighted by the relative strengths of the hyperfine components) as it appears in the CDMS catalogue. The rest frequency of the isolated hyperfine line, N$_2$H$^+$~(1,0,1-0,1,2), used to build the moment 1 and 2 images  is 93.1762595~GHz.}
    \label{tab:obslines}
\end{table*}

Outflows play an important role in the chemical and physical properties of the star-forming regions. They can induce changes in the chemical composition of their host clouds by dragging molecules formed in the disk or stellar atmosphere \citep[e.g.,][]{Bachiller1996}, and by producing new molecules as a result of collisions between the outflow with the cloud, or between outflows themselves (e.g., \citealt{Zapata2018,Toledano2023}). 
Additionally, outflows contribute with enough energy and momentum to the
maintain the turbulence \citep[e.g., ][]{Plunkett2015}, even though outflow feedback has a minor contribution compared to the
gravitational potential energy of the clump.

Moreover, while it is customary to see shocks produced in the internal working surfaces of protostellar outflows due to differences in --for instance-- the ejection velocity, shocks can also occur when outflows collide with dense obstacles such as protostellar cores, a dense pocket of gas, the walls of molecular clouds, or even when the outflows collide with each other \citep[e.g.,][]{Hartigan1987,Raga2002,Zapata2018}. When the obstacle encountered is smaller than the cross-section of the outflow, it becomes entrained, and a reverse bow shock is formed \citep[a so-called \lq\lq cloudlet shock\rq\rq; e.g.,][]{Hartigan1987,Hartigan2000,Raga1994}. Alternatively, when the obstacle is larger, the outflow could be deflected, forming a quasi-stationary shock at the point of collision followed by a spray of shocked outflow material downstream \citep[e.g.,][]{Raga2002}. Despite the ubiquity of jets and outflows in the Galaxy, only a few have been undoubtedly reported as being deflected or colliding \citep[e.g.,][]{Zapata2018,2022Lopez}. 

The high-mass star forming region GGD 27 is located in the dark molecular cloud L291 in Sagittarius arm at a distance of 1.4 kpc \citep[][]{Anez2020}. The (sub)millimeter emission in this region is dominated by two protostars MM1 and MM2 (\citealt{Girart2018}). The MM1 massive protostar drives the spectacular radio jet associated with HH 80-81 object \citep[e.g.,][]{Rodriguez1989,Marti1993,Carrasco2012,Masque2012,Masque2015}. This study focuses on MM2, which comprises two massive dust cores and disks --MM2(E) and MM2(W), also called ALMA 17 and ALMA 13-- apparently in a very early stage of evolution with masses of $1.9\pm0.2$ and $1.7\pm0.3$ \msun, respectively \citep{FernandezLopez2011a,FernandezLopez2011b}. \cite{Busquet2019} estimated that the lower limit of their disk masses to be 0.2 and 0.02 \msun, and their radii 120 and 30 \au, respectively. MM2(E) seems to be powering at least two molecular outflows: a SiO monopolar outflow due northeast and a CO bipolar precessing outflow oriented in a north-south direction \citep{FernandezLopez2013,Bally2023}. \cite{Qiu2009} through the line emission from high-density tracers such as CH$_3$OH, CH$_3$CN, H$_2$$^{13}$CO, OCS, and HNCO reported the existence of the source MC located at $3\arcsec$ north of MM2(E), located along the path of the northern lobe of the latter outflow. This source has no (sub)millimeter and/or infrared counterpart, but presents a very rich shock-like chemistry, which may be indicative of a stationary shock between the outflow and a dense pocket of molecular gas \citep{FernandezLopez2011b}. There is still another faint and compact dust core about $5\arcsec$ north from MM2(E). Finally, \cite{Kurtz2005} detected the emission from H$_2$O masers at the position of MM2(E), displaying a very broad range of blue-shifted velocities \citep[reaching $-100$ \kms, see also][]{Marti1999}. Also, \cite{Kurtz2004} reported a class I CH$_3$OH maser spot about $1\arcsec$ south of the MC source, in the range $+12$ to $+16$ \kms. 

This paper focuses on the molecular emission and kinematics associated with the northern redshifted lobe of the outflow from MM2(E). The paper is organized in the following way: Section \ref{sec:observations} details the observations. The results are presented in Section \ref{sec:results} and analyzed and discussed in Section \ref{sec:discussion}. Finally, the conclusions are shown in Section \ref{sec:conclusions}.

\section{Observations}
\label{sec:observations}

\subsection{3~mm Observations}
\label{subsec:obs3mm}

Observations of the GGD\,27 system were carried out with the Atacama Large Millimeter/submillimeter Array (ALMA) in Band\,3 on 2014 June 27 as part of the program 2012.1.00441.S (PI: Roberto Galv\'an-Madrid) during the Early Science Cycle 1 phase. In this work we focus on the emission of the hyperfine structure of the N$_2$H$^+$~(1--0) and the CH$_3$CN~(5--4) K-ladder lines. The calibrated data delivered by the ALMA European Regional Center were further self-calibrated and cleaned using Common Astronomy Software Application (CASA) package (\citealt{CASAteam}) version 6.4.0-16. A robust factor of 0.5 was applied to balance sensitivity and angular resolution and generate the molecular line cubes. Details of the rest frequencies, synthesized beam, spectral channel, and rms noise level are provided in Table \ref{tab:obslines}.

\subsection{1~mm Observations}
\label{subsec:obs1mm}

ALMA observations were obtained in Band\,7 on 2016 September 6 and 7 as part of the program 2015.1.00480.S (PI: Josep Miquel Girart) during Cycle 3 phase. The observations were reported in detail in \citet{FernandezLopez2023}. The H$_2$CO ($4_{1,3}-3_{1,2}$) line was included in the spectral window 27. This molecular transition was the only one in the spectral setup displaying significant emission associated to the main outflows originated by MM2(E). The data were calibrated using the CASA package version 4.7. Standard calibration and self-calibration using the continuum emission were applied to the spectral data. A robust factor of 0.5 to weigh the visibilities was applied to generate the H$_2$CO ($4_{1,3}-3_{1,2}$) channel maps. Details of the observations of this line are presented in Table \ref{tab:obslines}.

\section{Results}
\label{sec:results}

\begin{figure*}[t!]
\centering
\includegraphics[width=\linewidth]{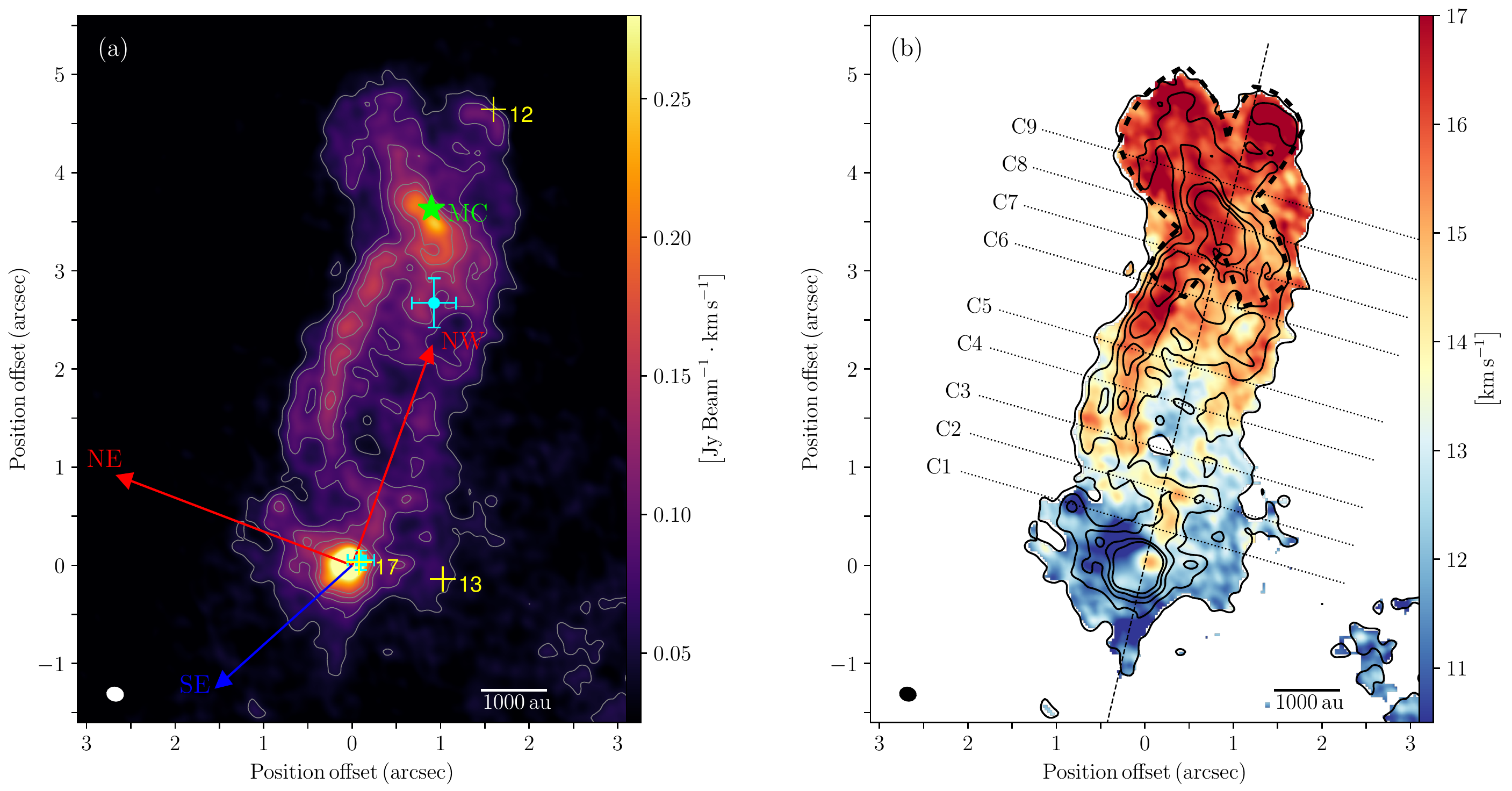}
\caption{GGD 27--MM2(E) molecular line emission of H$_2$CO. (a) ALMA moment zero or integrated intensity map. (b) ALMA first moment or intensity--weighted velocity map. The two red arrows represent the redshifted NE and NW CO outflow directions, while the blue arrow is the blueshifted SE CO outflow direction \citet{FernandezLopez2013}. Yellow crosses represent the position of the continuum sources reported by \citet{Busquet2019}, where source 13 and 17 are MM2(W) and MM2(E), respectively. Cyan dot and square mark the position of the CH$_3$OH and H$_2$O masers sources obtained from \citet{Kurtz2004} and \citet{Kurtz2005}, respectively. Green star represents the position of the source MC taken from \citet{Qiu2009}. The black dashed X-shaped curve and the black dashed line in (b) outlines the shocked region around the MC and the outflow axis, while the dotted lines C1--C9 indicate the position where the position-velocity of Figure \ref{fig:horizontal_pv_diagrams} have been extracted. The synthesized beam is shown in the lower left corner. The contour levels in both panels start at 5$\sigma$ in steps of 3$\sigma$, 6$\sigma$, 9$\sigma$, and 12$\sigma$, where $\sigma=9.5$\mjy\,\kms.}
\label{fig:moments}
\end{figure*}

\subsection{Morphology and kinematics of the outflow from MM2(E)}
\label{subsec:morphology}

Figure \ref{fig:moments} displays the emission of the H$_2$CO molecular line associated with the source GGD 27-MM2(E). The ALMA moment zero or integrated intensity map is shown in Figure \ref{fig:moments}a, while the ALMA first moment or intensity-weighted velocity map is presented in Figure \ref{fig:moments}b. The maps are integrated over a velocity range of 4-21 \kms (the systemic velocity of the MM2(E) is $\sim$11.5 \kms). The emission of the H$_2$CO extends up to $\sim 4^{\prime\prime}.8$ (6720~au) northwest of the main source (P.A.$\approx-14^{\circ}.5$), and it could be associated with the inner part of the CO molecular outflow previously reported by \citet{FernandezLopez2013}. 
Given that the observed H$_2$CO line is a relatively dense gas tracer \citep[n$_{crit}\approx10^6$ for a 50~K temperature, ][]{Shirley2015},
its overall spatial distribution, and the range of radial velocities displayed, we interpret that the emission in Figure \ref{fig:moments}a could be associated with the densest layers of the molecular outflow. 
The H$_2$CO observations also display weak emission of the southeastern outflow lobe \citep{Qiu2009,FernandezLopez2013} at line-of-sight velocities 8 to 12 \kms, close to the protostellar envelope, probably tracing the outflow cavity walls (see Figures \ref{fig:bluechannels} and \ref{fig:redchannels} on Appendix \ref{app:channel}). 

An interesting feature regarding the main body of the outflow is displayed in the velocity map (Figure \ref{fig:moments}b). It shows an overall increase of the line-of-sight velocity of the gas with the distance from the central source, presenting a clear south--north velocity gradient. This gradient may be explained as an outflow driven by a wide--angle wind \citep{Shu1991,Lee2000}, or a jet--driven wind \citep{Lee2001}, but also by a bulk outflow oscillation crossing the plane of the sky. Note that the velocity range presented in the velocity map (blue to red colors in the image) only covers the range of the redshifted emission with respect to the systemic velocity.

Downstream the outflow, there is a distinct X-shape observed in the north part as identified by a black dashed line in Figure \ref{fig:moments}b and channel maps of Figure \ref{fig:redchannels} of Appendix \ref{app:channel}. In this work we show evidence suggesting that this feature is associated with a collision between the outflowing gas from MM2(E) and the MC cloudlet (green star in Figure \ref{fig:moments}a). The collision between an outflow and a dense core has been detected in other sources such as IRAS~21391+5802 \citep{Beltran2002} and HH\,110-HH\,270 \citep{Kajdic2012}.

\subsection{Morphologhy in the MM2(E) Surroundings}
\label{subsec:structure}

\begin{figure*}[t!]
\centering
\includegraphics[scale=0.60]{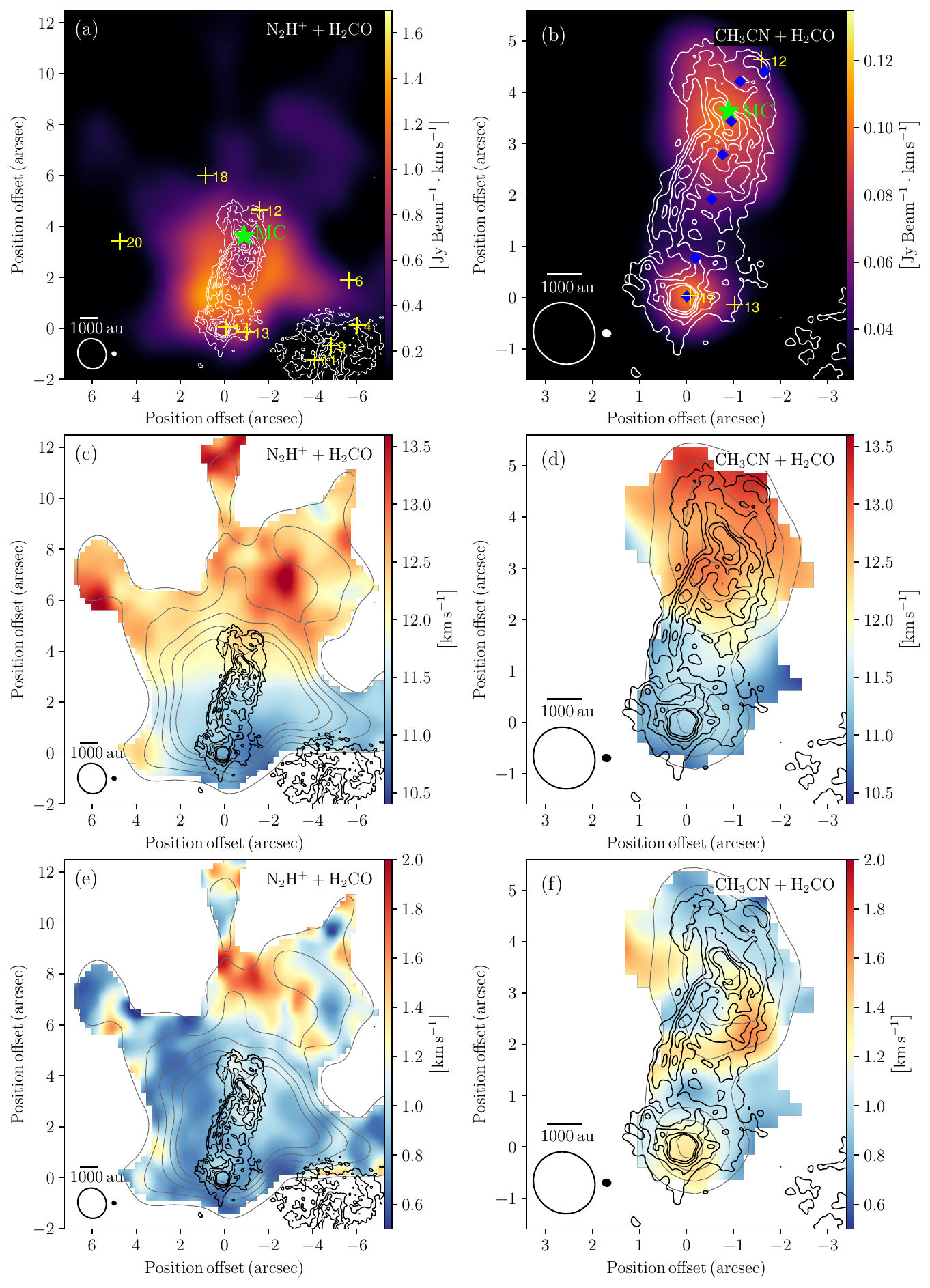}
\caption{GGD 27-MM2 molecular line emission from $\mathrm{N_2 H}^+$, $\mathrm{CH_3 CN}$, and H$_2$CO molecular lines. (a) Intensity map of the $\mathrm{N_2 H}^+$. (b) Intensity map of the $\mathrm{CH_3 CN}$. (c) Velocity map of the $\mathrm{N_2 H}^+$. (d) Velocity map of $\mathrm{CH_3 CN}$. (e) Velocity dispersion map of the $\mathrm{N_2 H}^+$. (f) Velocity dispersion map of the $\mathrm{CH_3 CN}$. The H$_2$CO molecular line emission is depicted in all panels in white (panels a and b) and black (panels c-f) contours, the contours levels are the same of Figure \ref{fig:moments}. The gray contours are the intensity map of the emission of $\mathrm{N_2 H}^+$ (panels c and e) and $\mathrm{CH_3 CN}$ (panels d and f), the contours levels start from 5$\sigma$ with steps of 5$\sigma$, 10$\sigma$, 15$\sigma$, and 20$\sigma$, where $\sigma=30.3$ \mjy\,\kms for the $\mathrm{N_2 H}^+$ emission, and start from 5$\sigma$ with steps of 3$\sigma$, 6$\sigma$, 9$\sigma$, and 12$\sigma$, where $\sigma=6.8$ \mjy\,\kms for the $\mathrm{CH_3 CN}$ emission. The synthesized beams are shown in the lower left corners. In panels a and b, green star presents the position of the source MC taken from \citet{Qiu2009}, and yellow crosses marks the positions of the continuum sources reported by \citet{Busquet2019}. Blue diamonds in panel b indicate the position where we have extracted the spectra of the $\mathrm{CH_3 CN}$ (see text).}
\label{fig:mosaic}
\end{figure*}

To test that the molecular outflow is colliding with the MC cloudlet, we compare the outflow structure with the emission of $\mathrm{N_2 H}^+$ and $\mathrm{CH_3 CN}$. Figures \ref{fig:mosaic}a and \ref{fig:mosaic}b present the intensity maps of $\mathrm{N_2 H}^+$ and $\mathrm{CH_3 CN}$ overlaid in white contours the emission of the H$_2$CO, respectively. The maps are integrated over a velocity range of -0.05--21.95 \kms\,and 9.37--22.10 \kms\,for the $\mathrm{N_2 H}^+$ and $\mathrm{CH_3 CN}$, respectively. The emission of the $\mathrm{N_2 H}^+$ traces the molecular cloud around the emission of the outflow, where protostars are forming as indicated by the continuum emission of these sources (yellow crosses), \citet{Busquet2019}. On the other hand, the $\mathrm{CH_3 CN}$ traces the emission around the central source MM2(E) and the MC cloudlet. 

The emission of the $\mathrm{N_2 H}^+$ traces the projected area of $\sim 8^{\prime\prime}\times7^{\prime\prime}$ ($\sim 11200$ \au $\times9800$ \au). This area corresponds to the emission of NH$_3$ previously reported by \citet{Gomez2003}, where they found that it corresponds to the molecular cloud around the protostellar system GGD 27-MM2, with a gas temperature of $\sim$ 150 K and a mass of H$_2$ of 16 \msun. It is noteworthy that the $\mathrm{N_2 H}^+$ has an internal cavity which spatially corresponds to the emission of the molecular outflow traced by H$_2$CO, therefore, we can assume that the molecular outflow has sweep up the material of the molecular gas.

As mentioned above, looks like the emission of the $\mathrm{CH_3 CN}$ is concentrated in the regions around the central source MM2(E) and MC. However, we detect a weaker emission of $\mathrm{CH_3 CN}$ between these regions, therefore, we may assume that the $\mathrm{CH_3 CN}$ also traces the inner part of the molecular outflow.

\subsection{Kinematics in the MM2(E) Surroundings}
\label{subsec:kinematics}

Figures \ref{fig:mosaic}c and \ref{fig:mosaic}e show the ALMA first moment or intensity-weighted velocity map and the ALMA second moment or intensity-weighted velocity dispersion map of the emission of $\mathrm{N_2 H}^+$, respectively. In both the velocity map and the velocity dispersion map, the gray and black contours are the $\mathrm{N_2 H}^+$ and H$_2$CO total intensity maps. These maps were integrated over a velocity range of -0.05--7.02 \kms, corresponding to hyperfine line $F_{1},F=1,0\to1,1$ \citep[e.g.,][]{Caselli1995,AlvarezGutierrez2024}. The difference in the integration range between the velocity maps and the intensity map arises because the $\mathrm{N_2 H}^+$ has seven hyperfine lines, of which only the first is not contaminated by the others. 

It is noteworthy that the cloud exhibits a south-north velocity gradient of $\sim 3$~\kms and a line width of $\sim1.5$~\kms. While the velocity gradient may be associated with the expansion of an irregular cloud, the line width could be produced by turbulence of the gas at different spatial scales.


Figures \ref{fig:mosaic}d and \ref{fig:mosaic}f depict the velocity and velocity dispersion maps of the emission of $\mathrm{CH_3 CN}$ with the intensity maps of $\mathrm{CH_3 CN}$ and H$_2$CO overlaid in gray and black contours, respectively. The velocity maps were integrated over a velocity range from 8.57--14.14 \kms, corresponding to hyperfine line K=0 \citep[e.g.,][]{Purcell2006}. Similar to the emission of $\mathrm{N_2 H}^+$, the integration range differs from that of the intensity map because the $\mathrm{CH_3 CN}$ has four distinct transitions (see \S\ref{subsec:obs3mm}), of which, three transitions are blended.

\begin{figure*}[t!]
\centering
\includegraphics[width=\linewidth]{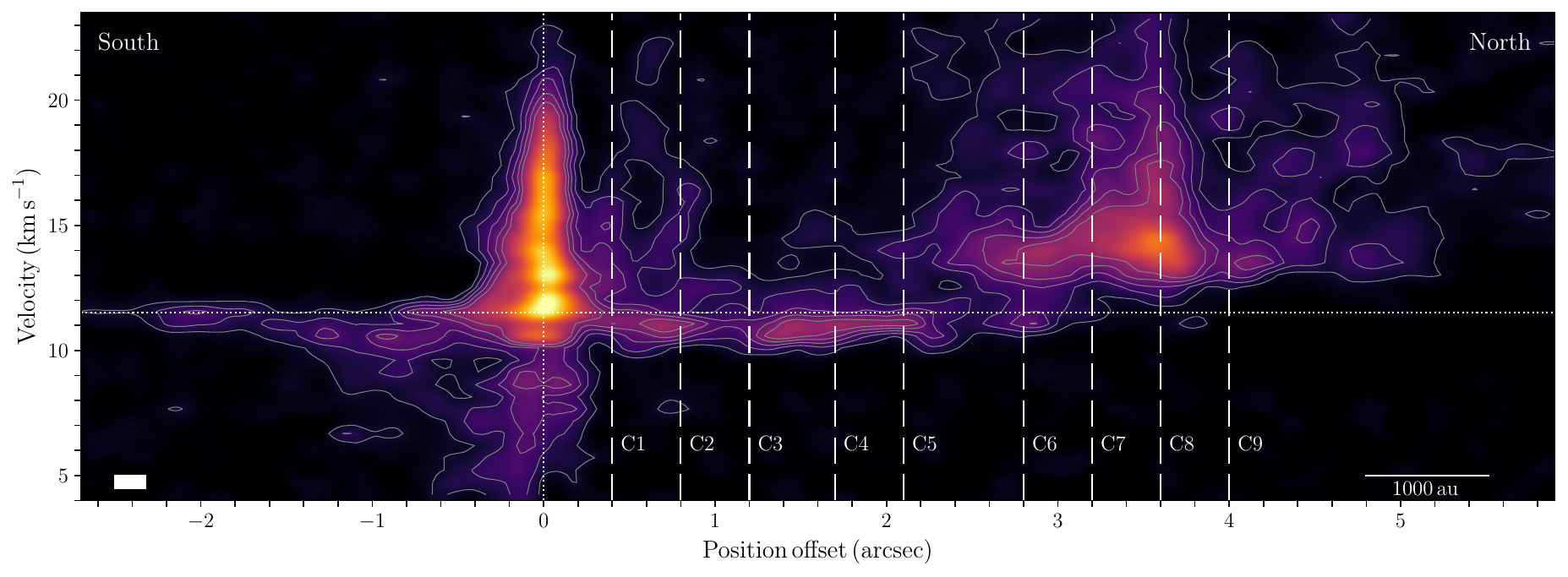}
\caption{Position--velocity diagram of the H$_2$CO emission along the outflow axis. This orientation of the axis is estimated by joining the peak of the MM2(E) continuum emission and the position of the MC source. Dotted lines are the position of the source (vertical) and the velocity of the system $V_\mathrm{sys}\simeq 11.5$ \kms (horizontal). Vertical dashed lines indicate the position where the perpendicular position-velocity diagrams are made (see Figure \ref{fig:horizontal_pv_diagrams}). The contour levels start at 2$\sigma$ in steps of 2$\sigma$, 4$\sigma$, 6$\sigma$, 8$\sigma$, where $\sigma=2.1$ mJy beam$^{-1}$. The white bar represents the angular resolution ($0^{\prime\prime}.18$) and the channel width (0.5 \kms).}
\label{fig:vertical_pv_diagram}
\end{figure*}

$\mathrm{CH_3 CN}$, like H$_2$CO, traces the molecular outflow and thus exhibits the same south-north velocity gradient. Nevertheless, the velocity gradient associated with $\mathrm{CH_3 CN}$ is of the order to $\sim$3 \kms, while the velocity gradient associated with H$_2$CO is $\sim$7 \kms. Additionally, the velocity dispersion map presents a line width of $\sim$1.5 \kms, which could be associated with the material being dragged or with shocks between the outflow and the cavity walls.

The emission of the H$_2$CO and $\mathrm{CH_3 CN}$ traces the molecular outflow associated with MM2(E) inside the cavity detected in the $\mathrm{N_2 H}^+$ cloud. Both molecules show similar kinematic features, and the outflow extends up to $\sim$$4^{\prime\prime}$, where it appears to be colliding with MC. In the following sections, we focus on the shocked region associated with the MC cloudlet.

\begin{figure*}[t!]
\centering
\includegraphics[width=\linewidth]{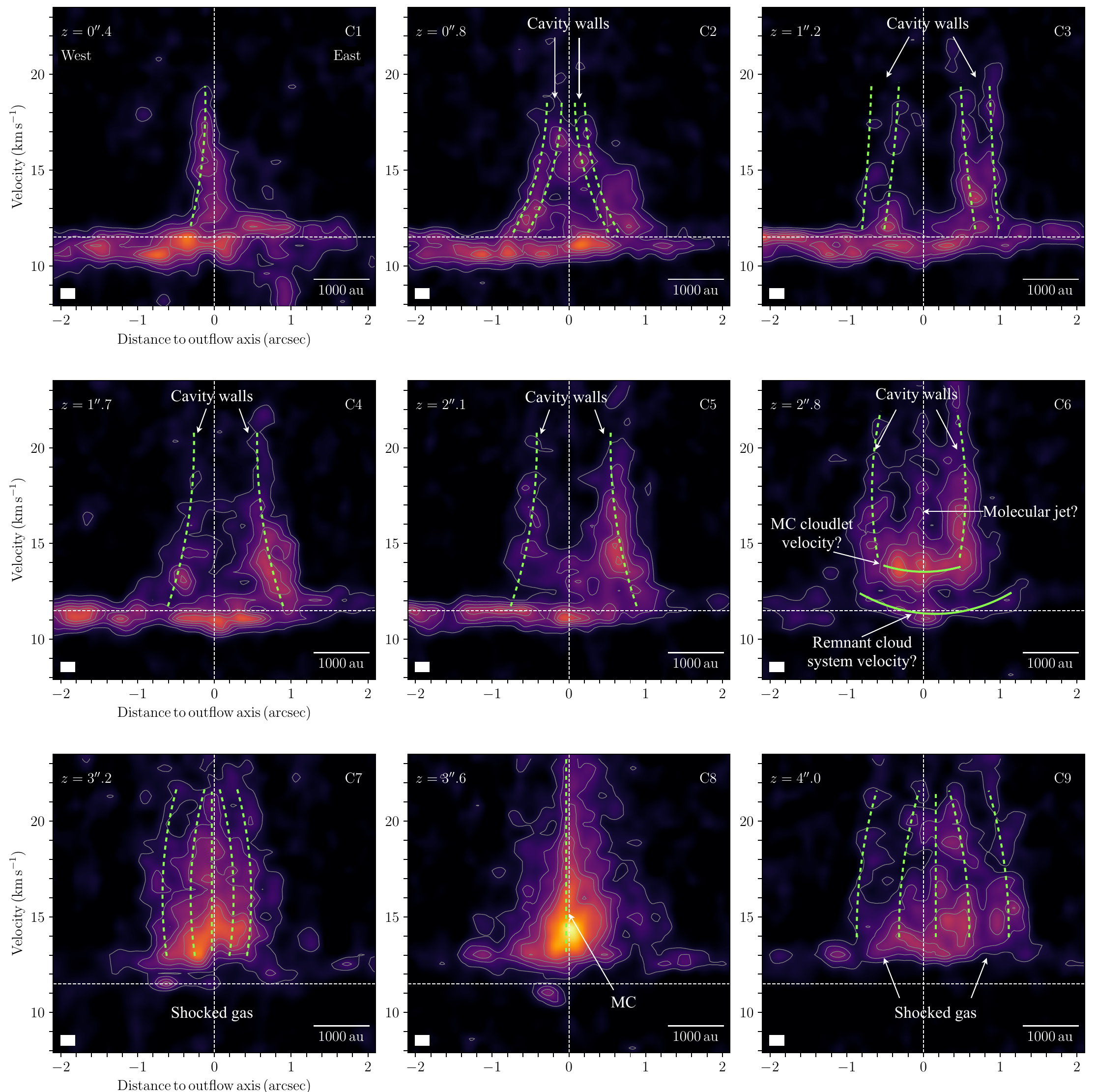}
\caption{Position-velocity diagrams perpendicular to the outflow axis of the H$_2$CO emission at different heights indicated by vertical dashed lines in Figure \ref{fig:vertical_pv_diagram}. White dotted lines represent the position of the outflow axis (vertical) and the systemic velocity at 11.5 \kms (horizontal). Green dashed lines in all panels trace the putative structures associated with the outflow cavity walls and shocked gas. The contour levels start at 2$\sigma$ in steps of 2$\sigma$, 4$\sigma$, 6$\sigma$, 8$\sigma$, where $\sigma=2.1$ mJy beam$^{-1}$. The white bar represents the angular resolution (0.$^{\prime\prime}$18) and the channel width (0.5 \kms).}
\label{fig:horizontal_pv_diagrams}
\end{figure*}

\subsection{Shocked region and molecular outflow}
\label{subsec:shockedgas}

Figure \ref{fig:vertical_pv_diagram} shows a Position--Velocity (PV) diagram of H$_2$CO emission created from a cut taken along the outflow axis (black dashed line in Figure \ref{fig:moments}b). The strongest emission presents a large velocity dispersion at the position of the protostellar disk/envelope (zero positional offset), which may be due to gas rotation around the protostar(s). The extended emission observed around the MM2(E) systemic velocity $\sim11.5$ \kms is associated with the molecular cloud or envelope mainly surrounding the lower half (up to $\sim3\arcsec$) of the outflow from MM2(E). A strong shock reaching radial velocities over 22\,\kms is present at $\sim3\farcs6$. However, the high-velocity shocked gas extends from $2\farcs8$ to $5\farcs1$ from the protostar. 
In general, we note that the PV diagram does not present the characteristic parabolic shape expected for a wide--angle disk wind \citep{Lee2000}, but a convex spur-like structure, similar to the shape expected for a jet--driven outflow inclined with respect to the plane of the sky \citep[see e.g., Figure 5 of][]{Lee2001}.

\begin{figure*}[t!]
\centering
\includegraphics[width=\linewidth]{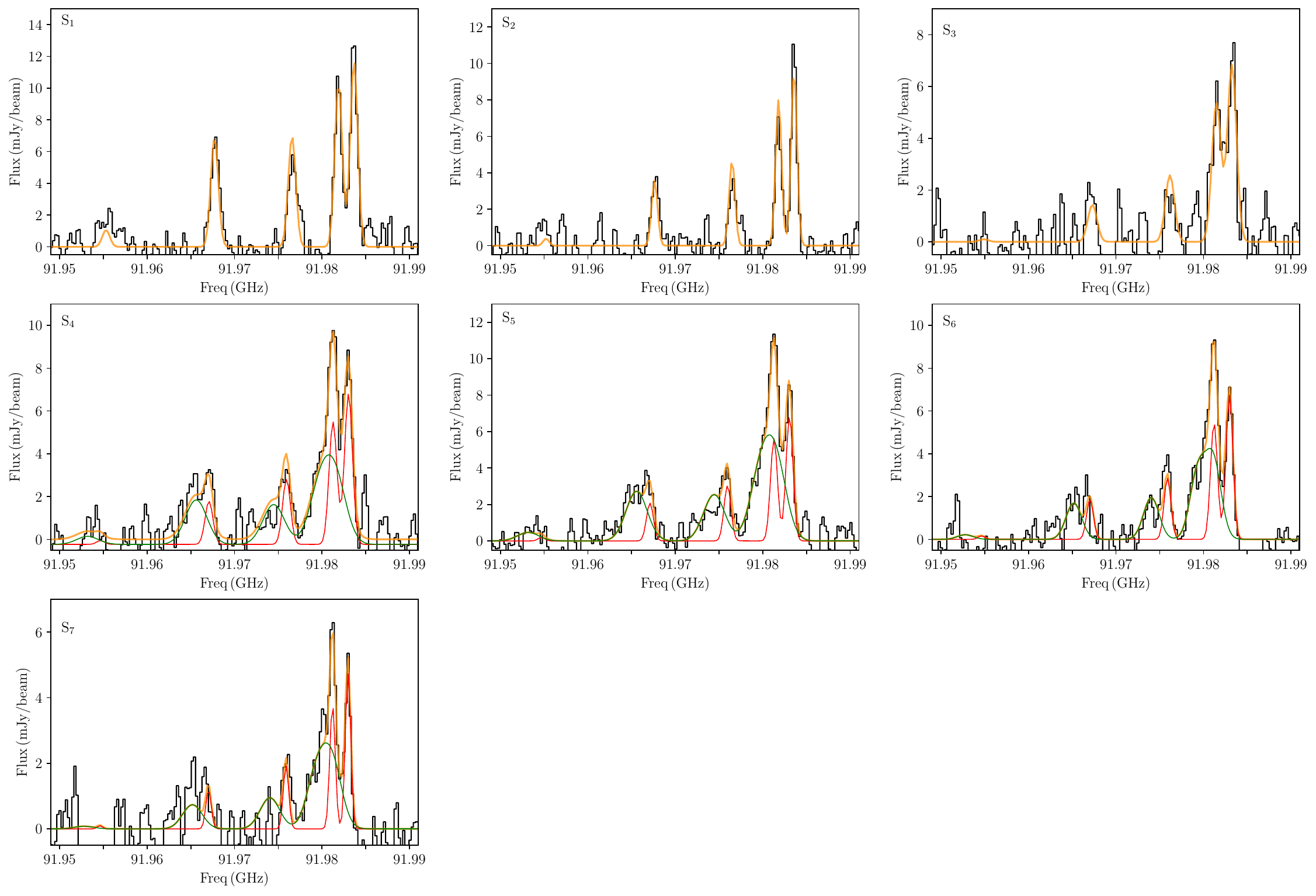}
\caption{Extracted spectra of the CH$_3$CN (5-4) K-ladder from the blue dots in panel b of Figure \ref{fig:mosaic}. The labels S1 to S7 indicate the location of the points, with S1 being the point over the MM2(E) and S7 the most distant point in the south-north direction. The black line represents the data obtained from the observations, while the orange line shows the best fit obtained. For panels S1 to S3, we use only one velocity component for our fit, whereas for panels S4 to S7, the addition of two velocity components (green and red lines) was utilized.}
\label{fig:spectra}
\end{figure*}

The vertical dashed lines (C1--C9) in Figure \ref{fig:vertical_pv_diagram} indicate the positional offsets used to generate a series of PV diagrams perpendicular to the outflow axis (Figure \ref{fig:horizontal_pv_diagrams}). Analyzing these cuts in detail, we find that in the region closer to the MM2(E) disk/envelope (panels C1--C3), the PV diagrams trace the emission associated with the cavity walls of the molecular outflow. The walls are identified as almost vertical finger-like structures with large spreads in velocity (similar to shocks) or elongated blobs at high velocities. In particular, panel C1 presents an unresolved structure with a high-velocity dispersion (green dashed line); this spatially unresolved structure could be regarded as material directly ejected from the disk and/or as the outflow cavity walls \citep[formed as a result of the interaction between the outflow and the ambient medium,][]{Lee2000}. Panel C2 shows the emission of the cavity walls of the molecular outflow (external dashed lines), these walls appear as elongated fingers with a high velocity dispersion ($\sim$6 \kms). The unresolved emission region between the external cavity walls observed in panel C2 could be associated with an internal layer of the molecular outflow, which has been previously detected in several sources, such as HH 46/47 (\citealt{Zhang2019}), DO Tauri (\citealt{FernandezLopez2020}), DG Tau B (\citealt{dValon2020,dValon2022}), and HH 30 (\citealt{JALV2024}). This internal layer is confirmed in panel C3, where the emission of the four elongated fingers with similar velocity dispersion ($\sim$9 \kms) is observed. In this panel, we interpret the cavity walls as being traced by the external fingers (external dashed lines), located at $\sim\pm 1^{\prime\prime}$ of the outflow axis, while the internal layer is associated with the internal fingers (internal dashed lines), located at $\sim\pm0.6^{\prime\prime}$ from the outflow axis.

In the central region (panels C4--C5), the H$_2$CO emission traces two different finger-like structures associated with the cavity walls. This same pattern has been observed in other sources where the inclination angle with respect to the plane of the sky is $i\geq 30^{\circ}$ such as HH 46/47 (\citealt{Zhang2019}) and DG Tau B \citep[][]{dValon2020,dValon2022}. 
The H$_2$CO is also tracing the extended cloud emission comprising all the sources in this region \citep[see channels from 10.0 to 11.5 \kms and][]{FernandezLopez2011b}, which indicates that the observed emission from the walls probably comes from gas dragged by the outflow. 
A distinct aspect, evident also from the moment 0 map (Figure \ref{fig:moments}), is that it has a brighter eastern side in H$_2$CO. 
This feature, along with the characteristic shape of MM2(E)'s molecular outflow, could be associated with the presence of a very asymmetric jet such as L1157 \citep{Podio2016}, or with two or more episodic ejections of material as observed in IRAS 21078+5211 \citep{Moscadelli2022,Moscadelli2023}. An alternative explanation is that the density spatial distribution of the surrounding material is not homogeneous, which may cause differences in the excitation of molecules and produce asymmetries in the outflow shape emission.

In regions far away from the central source (panels C6--C9), the velocity cuts mainly show the formation of a strong shock and the downstream perturbations of the gas produced by the putative collision between the outflow and the MC cloudlet. Panels C6 and C7 show the behaviour of the gas just upstream the MC position (C8 dashed line in Figure \ref{fig:vertical_pv_diagram}). 
Both panels show the walls of the molecular outflow, and possibly the molecular jet impacting dense gas at a more redshifted velocity than the systemic velocity ($\approx$13--14 \kms) and creating a shock. The horizontal structures observed in panel C6 at $\sim$11.5 \kms~ and $\sim$14 \kms~ could be associatiated with the remnant cloud system velocity of the MM2(E) and with the MC cloudlet, respectively.
Panel C9 traces shocked material in a more turbulent and spray-like way. The measured outflow width downstream MC is larger ($\sim$1$\farcs0$) than the outflow region from the main body (between $1\farcs7$ and $2\farcs8$; panels C4-C6), which is $\sim$0$\farcs6$. 
This width increase also suggests and supports the scenario of an upstream pounding of the outflow with a dense cloudlet \citep{Raga2002}. 

\section{Analysis and Discussion}
\label{sec:discussion}
\subsection{Outflow mass and energy}
\label{subsec:massenergy}

Here we make an estimation of the column density of the outflow gas using the expressions from \cite{Mangum2015} adapted to the H$_2$CO ($4_{1,3}-3_{1,2}$) line emission. The H$_2$CO ($4_{1,3}-3_{1,2}$) line emission considered (with an intensity over $3-\sigma$) is generally optically thin through MM2(E)'s outflow, so we use the simplified expression of equation 86 in \citet{Mangum2015}. As a proxy for the excitation temperature we utilize a range of values between 30~K and 70~K derived in the analysis of the CH$_3$CN K-ladder emission (see Section \ref{sec:shocks}). From this, we also derive the mass, momentum, and kinetic energy of the outflow following an analogous procedure as, e.g., in \cite{Vazzano2021} (note that we give the momentum and energy without correcting by the unknown outflow inclination with respect to the line of sight).  We adopt an H$_2$CO abundance with respect to H$_2$ of $10^{-10}$ \citep{Gerner2014,Tang2018,Gieser2021,Zhao2024}, which produces an outflow mass between 1.7 \msun\,and 2.1 \msun. Likewise, the derived momentum, $P_{outflow}$, and kinetic energy, $E_{outflow}$, ranges are 7.8--10.1 \msun\,\kms and 5.0--6.6$\times10^{44}$~erg, respectively. 

The size of the molecular outflow is $z\simeq 4\farcs8\simeq6720\,\mathrm{au}$, while its outward velocity is $V_z\simeq 9\,\mathrm{km\,s^{-1}}$ (this velocity is the difference between the maximum observed velocity in channel maps in Figure \ref{fig:redchannels}, $\sim 21\,\mathrm{km\,s^{-1}}$, and the systemic velocity $\sim 11.5\mathrm{km\,s^{-1}}$), therefore we obtain the dynamical time $\tau_{dyn}=z/V_z\simeq 3500$ yr.
We estimate a mass-loss outflow rate of $\dot{M}_{outflow}=$4.9--6.0$\times10^{-4}$ \msun\,yr$^{-1}$. These values are consistent with the previously reported of outflows driven by massive stars \citep[e.g., ][]{LopezSepulcre2009,SanchezMonge2013,Maud2015}.

\begin{figure*}[t!]
\centering
\includegraphics[width=\linewidth]{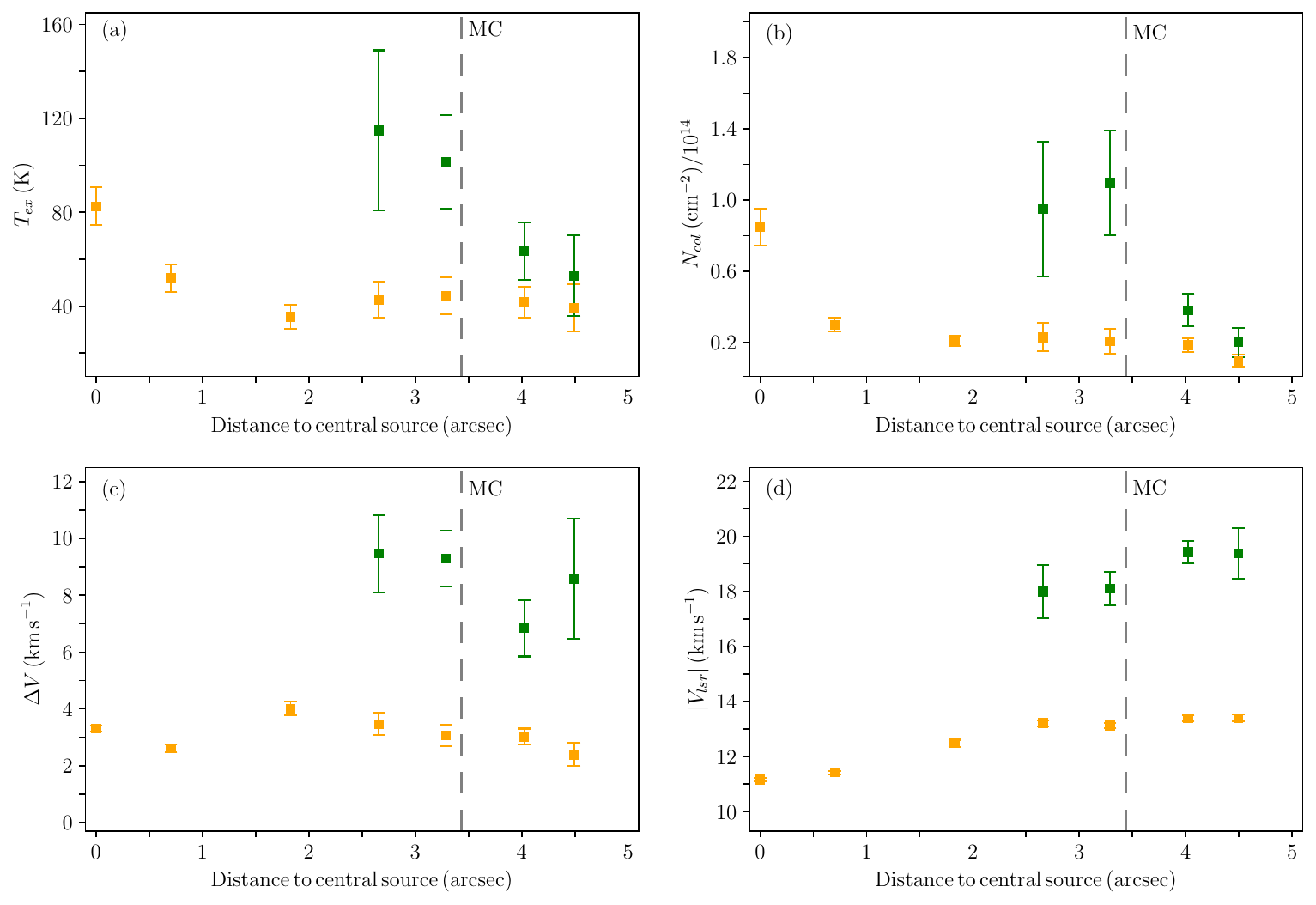}
\caption{Physical conditions of the gas of the emission of $\mathrm{CH_3 CN}$ associated with the GGD 27-MM2(E) molecular outflow as a function of the distance to central source. (a) Excitation temperature. (b) Column density. (c) Line width. (d) Velocity of the gas. Error bars are derived from the model (see text). Orange and green squares result from a  simultaneous fitting of two  kinematic components. The orange squares are probably associated with weak/mild shocks from the expanding outflow walls, whereas the green squares are possible linked with a more violent shock between the fast wind and a dense cloudlet (MC source), which position is marked by a  dashed vertical line at about $3\farcs5$.} 
\label{fig:parameters}
\end{figure*}

We next derive the mass of the dense gas clump, part of the parental molecular cloud enveloping the outflow from MM2(E). For this we use the HfS fitting tool on the hyperfine structure of the N$_2$H$^+$ (1-0) transition \citep[][panel a in Figure \ref{fig:mosaic};]{2017Estalella}. We estimate the average opacity of the main hyperfine line is about 0.3$\pm0.1$, which is taken into account to estimate the column density using the spectrum extracted over an area of 73 square arcseconds, limited by the 10$\sigma$ contour in the integrated intensity map (velocity interval used: 9.9--15.3 \kms; see panel c in Figure  \ref{fig:mosaic}), and the optically thin approximation \citep[equations 80 and 86 in ][]{Mangum2015}. We also found an excitation temperature of 23$\pm4$~K, which agrees well with the value measured by \cite{Gomez2003} for the ammonia temperature of the emission associated with this clump. We use a value for the partition function Q$_{rot}$ interpolating the values tabulated by the Cologne Database for Molecular Spectroscopy\footnote{https://cdms.astro.uni-koeln.de/classic/} with a power law of the form Q$_{rot}(T)\approx3.88T^{1.02}$, which nicely fits the data between 9~K and 500~K. The N$_2$H$^+$ column density is 1.4$\times10^{14}$~cm$^{-2}$, and adopting an N$_2$H$^+$ abundance ranging between 1-2$\times10^{-10}$ \citep{Shirley2005, AlvarezGutierrez2024}, we derive a hydrogen column density of 7.1--14.2$\times10^{23}$~cm$^{-2}$. The total mass of the whole clump of gas is therefore 13.3--26.5 \msun, a factor of 8 to 12 larger than the gas moved by the red-shifted lobe of MM2(E)'s outflow. This mass is, again, consistent with that estimated by \cite{Gomez2003}. 

With all of this we can address now the potential of the outflow from MM2(E) to introduce and maintain turbulence in its surroundings and whether the outflow can potentially overcome the cloud's gravity and disperse the gas. Following an analogous pathway as \cite{Plunkett2015} based on the \lq\lq outflow-regulated star formation\rq\rq\, model by \cite{Nakamura2014}, we first estimate the turbulent energy as $E_{turb}=\frac12 M_{cl}\sigma^2_{3D}$, with $\sigma_{3D}=(3/8ln(2))^{0.5}\Delta v_{turb}$. From the N$_2$H$^+$ emission, we estimate the dense part of the cloud surrounding MM2(E) has an approximate radius of $5\arcsec$ ($\sim7000$~au), and a $\sigma_{3D}=1.49$\kms, estimated by using $\Delta v_{turb}=\Delta v_{observed}-\Delta v_{thermal}$. The turbulent energy is $2.9-5.9\times10^{44}$~erg, comparable to MM2(E)'s outflow energy. In addition, would the cloud be spherical, its free-fall time is 2.0--2.8$\times10^4$~yr (for $M_{cl}$ values 13.3--26.5\msun). Hence, the free-fall time $t_{ff}$ is about a factor of 8 larger than the outflow dynamical time (we estimate $t_{dyn}\simeq3500$~yr for the outflow of MM2(E), although without correcting for its inclination). From $t_{ff}$ we can derive the momentum dissipation time, $t_{diss}=t_{ff}\cdot3.9~\kappa / M_{rms}\simeq1.7\times10^4$~yr, using that the parameter $\kappa\sim1$ (turbulence driving length approximately matches the Jean's length) and estimating the Mach number as $M_{rms}=\sigma_{3D}\cdot\sqrt{2.72~m_H/(K_B~T)}\simeq5.6$. 
Therefore, the momentum rate of the outflow ($\dot{P}_{outflow}=P_{outflow}/\tau_{dyn}\approx 1.4-1.8\times10^{28}$~erg~cm$^{-1}$) is about 7--9 times larger than the momentum dissipation rate of the internal turbulent motion in the cloud \citep[$\dot{P_{cl}}=-0.21 M_{cl} \sigma_{3D}/t_{diss}\approx -2\times10^{27}$~erg~cm$^{-1}$; see][]{Nakamura2014}, and it is plausible that the studied outflow is entering and replenishing a significant amount of turbulence in its parental cloud at spatial scales smaller than its approximate length ($\sim$7$\arcsec$), since it completely crosses its most dense part.

To compare the outflow energy and the gravitational binding of the cloud $W$, we use the expression $W=-GM_{cl}^2/R_{cl}$, where $G$ is the gravitational constant, $M_{cl}$ is the mass of the cloud, and $R_{cl}$ is the radius of the cloud. In this case, $W$ ranges from $-5.0\times10^{44}$~erg to $-1.8\times10^{45}$~erg (again depending on the mass of the cloud). \citet{Nakamura2014} defined the non-dimensional parameter $\eta_{out}=-2E_{outflow}/W$ to compare the cloud and the outflow energies. For MM2(E), $\eta_{out}$ ranges between 0.55 and 2.64. 
These relatively high values of $\eta_{out}$ indicate that MM2(E)'s outflow is very powerful, can disrupt/unbind large amounts of gas, and has the ability to erode large parts of its natal cloud. It has the potential to destroy or tremendously perturb smaller scale cores, such as ALMA~12's or the MC, directly impacting sites of ongoing star-formation. Even if we may have underestimated the abundance of H$_2$CO (which may result in a lower outflow mass and energy), and it is larger by a factor of 5, $\eta_{out}$ would be close to 1, indicating the power of this outflow.

\subsection{The shock at the position of the MC core}
\label{sec:shocks}

Figure \ref{fig:mosaic} shows a deep N$_2$H$^+$ cavity (panel a) and a CH$_3$CN emission enhancement toward the MC location (panel b). This spatial anti-correlation between N$_2$H$^+$ and CH$_3$CN emissions closely follows the MM2(E) outflow path.
The CH$_3$CN emission could be enhanced by the shocks of the outflow revealed by Figures \ref{fig:vertical_pv_diagram} and \ref{fig:horizontal_pv_diagrams}. At the same time, the physical conditions of the post-shocked gas may destroy the N$_2$H$^+$ from the quiescent gas of the molecular cloud \citep[the molecule is easily destroyed by CO released from the grain mantles when gas temperature increases over 20~K,][]{Jorgensen2004}, which may also be dragged and dispersed by the outflow. A decrease on the abundance of N$_2$H$^+$ has also been seen in shocks associated with other outflows \cite[][]{Tobin2012,Podio2014}. Moreover, the same anti-correlation between N$_2$H$^+$ and CH$_3$CN is seen at the position of MM2(E), where an increase in temperature and the presence of multiple outflows could be causing it.

The CH$_3$CN emission is seen at the location of MM2(E) and toward the northern lobe of its outflow, but it is specially enhanced in the area surrounding the MC position (about 4000~au$\times$3000~au in size). To check the physical conditions of the outflow gas we have extracted the spectra of the CH$_3$CN~(5--4) K--ladder from several locations along its main path (blue dots panel b in Figure \ref{fig:mosaic}). At every position we fitted synthetic spectra derived by solving the radiative transfer equation for an isothermal object in LTE \citep[following a procedure analogous to e.g.,][]{Moller2017}, using one or two independent layers when the spectra showed one or two kinematic components, respectively. The fitting was performed with a Python code which uses a nonlinear least-squares minimization algorithm to search for the best fit of the observed data \citep{Newville2020}. The results from the fits are presented in Figure \ref{fig:spectra}. In addition, Figure \ref{fig:parameters} presents a summary of the fitted parameters as a function of distance to the MM2(E) protostar. At any distance from the position of MM2(E), the CH$_3$CN shows supersonic line widths ($>2.4$ \kms), with excitation temperatures over $\approx40$~K. 
At distances from MM2(E) larger than $\sim2\arcsec$ two-velocity components (red and green lines in panels S4 to S7 of Figure \ref{fig:spectra}) are included in the fitting. Incorporating these components into positions S4 through S7 improves the model and reduces the residuals.
An inspection to Figure \ref{fig:parameters} shows a line component with a velocity between 11.0 and 13.5 \kms (orange squares in panel (d); see also the overall south-north velocity gradient in various gas tracers in Figures \ref{fig:mosaic}, \ref{fig:bluechannels} and \ref{fig:redchannels}), persistent beyond the $2\arcsec$. The gas traced by this component is warm (40--50~K), with mild line-widths (2--4 \kms), and a stable column density of about $2\times10^{13}$~cm$^{-2}$. At distances from MM2(E) larger than $2\arcsec$ there is a second high-velocity component (at $>18$ \kms), with warmer (T$_{ex}$ ranges 55--120~K) and higher density gas ($\sim10^{14}$~cm$^{-2}$). This component is probably related with strong red-shifted shocks ($\Delta V\approx8$ \kms). 

While the two kinematic CH$_3$CN components are likely due to shocks, the colder and less dense component seems to trace supersonic but mild shocks, maybe associated with the outflow cavity expansion against the parental cloud \citep[seen in N$_2$H$^+$ and NH$_3$,][]{Gomez2003}, and the warm and dense component could be linked to a more extreme interaction. To explain the latter, 
we consider mentioning here two alternatives:
(1) A collision between the outflow and a dense pocket of gas or cloudlet \citep[e.g.,][]{Raga1998,Hartigan2003}, or (2) the outflow is escaping the rear part of the molecular cloud, producing strong shocks \citep[see analogous cases in HH 80-81, HH46/47, and IRAS~16059-3857][]{Heathcote1998,Arce2013,Vazzano2021}. On one hand, the MC (and possibly ALMA~12 as well) position matches with shocked gas profiles \citep[large line-widths and spectral wings with high redshifted velocity; Figure \ref{fig:moments}, see also][]{FernandezLopez2011b}. This can be explained by the impinging fast wind which is ablating portions of the MC dense cloudlet (which was initially at a slightly redshifted radial velocity than MM2(E)).

On the other hand, the point at which the flow starts showing the two spectral components and the strong shock profiles coincides with an overall decrease of the N$_2$H$^+$ abundance that may be tracing the north edge of the denser part of the molecular cloud. This morphology suggests that fast, very hot bullets may expand violently as they escape from the cloud into a low pressure and hot surrounding, such as in HH~80A and HH~81A \citep{Heathcote1998}. In addition, the H$_2$CO flow emission does not extend further north, although there is evidence that the associated jet does \citep{Bally2023}.

Other piece of evidence that supports the presence of these extreme shocks is the H$_2$CO X-shape at the location of the old MC source (green star in Figures \ref{fig:moments} and \ref{fig:bluechannels}). Indeed, north of the crossing point at MC, the gas appears more turbulent, less ordered (see Figure \ref{fig:horizontal_pv_diagrams}), as expected after a collision of a fast wind an a dense cloudlet partially ablated \citep[see bottom panel in Figure 2 of ][]{Raga1998}). 

\subsection{Several outflow ejections?}
\label{subsec:streamlines}

In the region close to the central source, $z\leq 1^{\prime\prime}.2$, the PV diagrams depicted in Figure \ref{fig:horizontal_pv_diagrams} show (particularly in panel C3) that the emission of the H$_2$CO is dominated by two distinct layers: an external layer at a distance of $\sim$$\pm 1^{\prime\prime}$ from the outflow axis and an internal layer located at $\sim$$\pm0^{\prime\prime}.6$. These layers are also observed in the emission above 3$\sigma$ in the channel maps of Figure \ref{fig:redchannels}, particularly in the velocity range of $\sim$ 12.5--16 \kms. The multi-layer structure has been observed in several sources, such as HH 46/47 (\citealt{Zhang2019}), DO Tauri (\citealt{FernandezLopez2020}), DG Tau B (\citealt{dValon2020,dValon2022}), and HH 30 (\citealt{JALV2024}), where the authors suggested that the multiple layers are associated with episodic ejections of the material from the accretion disk. Assuming that the two layers correspond to different ejections, the outer layer could represent an older ejection, while the internal layer is related to a younger ejection. This implies that the material from the younger ejection has not traveled for a long time and is therefore not detected at heights $z>$$1^{\prime\prime}.2$.

Another possible explanation for the shape observed in panels C1--C3 of Figure \ref{fig:horizontal_pv_diagrams} and in the channel maps of Figure \ref{fig:bluechannels} is that the H$_2$CO emission is tracing multiple streamlines created by the presence of overdensities launched from the accretion rotating disk with an inclination from the polar direction \citep[e.g., IRAS~21078+5211][]{Moscadelli2023}. However, given that our observations do not have enough spatial resolution and sensitivity to resolve and trace the full path of all the possible streamlines, we cannot confirm this scenario.

\section{Conclusions}
\label{sec:conclusions}

We present a detailed analysis of ALMA observations of the molecular line emissions of the H$_2$CO, $\mathrm{CH_3 CN}$, and $\mathrm{N_2 H}^+$ from the molecular outflow, as well as the emission of the $\mathrm{N_2 H}^+$ from the molecular cloud, both associated with the protostellar system GGD~27-MM2(E). Our main results are the following.

\begin{enumerate}
    \item We detect that the emission of $\mathrm{N_2 H}^+$ traces a molecular cloud surrounding GGD27 MM2(E) and MM2(W) (ALMA 17 and ALMA 13, respectively). Additionally, other continuum sources such as ALMA 6, ALMA 12, and ALMA 18 are located within the cloud. The $\mathrm{N_2 H}^+$ cloud exhibits an internal cavity that spatially corresponds to the emission of the molecular outflow traced by H$_2$CO. We estimate that the mass of the envelope is between 13.3--26.5 \msun; this value is consistent with those previously reported by \citet{Gomez2003}.
    \item The molecular outflow within the cloud is depicted by the emission of H$_2$CO and $\mathrm{CH_3 CN}$. While the H$_2$CO is associated with the material dragged from the cloud and traces the walls of the outflow, the emission of $\mathrm{CH_3 CN}$ is concentrated around the central source MM2(E) and MM2(W), as well as in the vicinity of the MC. This distribution could indicate that the outflow is colliding with the MC.
    \item Through the emission of H$_2$CO, we estimate that the outflow has a mass between 1.7--2.1 \msun, a momentum of 7.8--10.1 \msun\,\kms, a kinetic energy of 5.0--6.6$\times$10$^{44}$ erg, which are consistent with outflows driven by massive stars. On the other hand the mass-loss rate of the outflow is 4.9--6.0$\times$10$^{-4}$ \msun\,\yr, consistent with the values previously reported. 
    \item We quantified the turbulence and gravity energies and found that the outflow contributes a significant amount of turbulence to its parent cloud. We find that the outflow is very powerful and has the ability to erode large parts of its parent cloud. It also has the potential to destroy or perturb smaller cores, such as ALMA 12 and MC.
    \item The X-shape feature detected to the north of the H$_2$CO molecular outflow suggests that this outflow could be colliding with the MC cloudlet. This collision is supported by the emission of $\mathrm{CH_3 CN}$, where we observe an increase in excitation temperature, column density, line width, and velocity of the gas in the region around the MC. This increase in all physical properties is associated with the perturbation of the gas due to the shock. The perturbation is also evident in the PV diagrams of the H$_2$CO emission, where we detect spray-shocked material downstream of the MC position with a larger width than the main outflow body.
\end{enumerate}

\begin{acknowledgements}
      We thank an anonymous referee for very useful suggestions that improved the presentation of this paper.
      We are indebted to Pablo Fonfr\'ia for his fruitful discussions and advices. J.A.L.V. and C.F.L. acknowledge grants from the National Science and Technology Council of Taiwan (NSTC 112–2112–M–001–039–MY3). M.F.L. acknowledges support from the European Research Executive Agency HORIZON-MSCA-2021-SE-01 Research and Innovation programme under the Marie Skłodowska-Curie grant agreement number 101086388 (LACEGAL). M.F.L. also acknowledges the warmth and hospitality of the ICE-UB group of star formation. S.C. acknowledges financial support from UNAM-PAPIIT IN107324, and CONACyT CF-2023-I-232 grants, M\'exico. J.M.G. and G.B. acknowledge support from the PID2020-117710GB-I00 and PID2023-146675NB-I00 grants funded by MCIN/ AEI /10.13039/501100011033. L.A.Z. acknowledges financial support from CONACyT-280775 and UNAM-PAPIIT IN110618, and IN112323 grants, M\'exico. R.G.M. acknowledges support from UNAM-PAPIIT project IN108822. This paper makes use of the following ALMA data: ADS/JAO.ALMA\# 2015.1.00480.S. ALMA is a partnership of ESO (representing its member states), NSF (USA) and NINS (Japan), together with NRC (Canada), NSTC and ASIAA (Taiwan), and KASI (Republic of Korea), in cooperation with the Republic of Chile. The Joint ALMA Observatory is operated by ESO, AUI/NRAO and NAOJ.
\end{acknowledgements}

%
%
\bibliographystyle{aa} 
\bibliography{sample631} 

\begin{thebibliography}{82}
\expandafter\ifx\csname natexlab\endcsname\relax\def\natexlab#1{#1}\fi

\bibitem[{{A{\~n}ez-L{\'o}pez} {et~al.}(2020){A{\~n}ez-L{\'o}pez}, {Osorio}, {Busquet}, {Girart}, {Mac{\'\i}as}, {Carrasco-Gonz{\'a}lez}, {Curiel}, {Estalella}, {Fern{\'a}ndez-L{\'o}pez}, {Galv{\'a}n-Madrid}, {Kwon}, \& {Torrelles}}]{Anez2020}
{A{\~n}ez-L{\'o}pez}, N., {Osorio}, M., {Busquet}, G., {et~al.} 2020, \apj, 888, 41

\bibitem[{{{\'A}lvarez-Guti{\'e}rrez} {et~al.}(2024){{\'A}lvarez-Guti{\'e}rrez}, {Stutz}, {Sandoval-Garrido}, {Louvet}, {Motte}, {Galv{\'a}n-Madrid}, {Cunningham}, {Sanhueza}, {Bonfand}, {Bontemps}, {Gusdorf}, {Ginsburg}, {Csengeri}, {Reyes}, {Salinas}, {Baug}, {Bronfman}, {Busquet}, {D{\'\i}az-Gonz{\'a}lez}, {Fernandez-Lopez}, {Guzm{\'a}n}, {Koley}, {Liu}, {Olguin}, {Valeille-Manet}, \& {Wyrowski}}]{AlvarezGutierrez2024}
{{\'A}lvarez-Guti{\'e}rrez}, R.~H., {Stutz}, A.~M., {Sandoval-Garrido}, N., {et~al.} 2024, \aap, 689, A74

\bibitem[{{Anglada} {et~al.}(2007){Anglada}, {L{\'o}pez}, {Estalella}, {Masegosa}, {Riera}, \& {Raga}}]{Anglada2007}
{Anglada}, G., {L{\'o}pez}, R., {Estalella}, R., {et~al.} 2007, \aj, 133, 2799

\bibitem[{{Arce} {et~al.}(2013){Arce}, {Mardones}, {Corder}, {Garay}, {Noriega-Crespo}, \& {Raga}}]{Arce2013}
{Arce}, H.~G., {Mardones}, D., {Corder}, S.~A., {et~al.} 2013, \apj, 774, 39

\bibitem[{{Bachiller}(1996)}]{Bachiller1996}
{Bachiller}, R. 1996, \araa, 34, 111

\bibitem[{{Bally}(2016)}]{Bally2016}
{Bally}, J. 2016, \araa, 54, 491

\bibitem[{{Bally} \& {Reipurth}(2023)}]{Bally2023}
{Bally}, J. \& {Reipurth}, B. 2023, \apj, 958, 99

\bibitem[{{Beltr{\'a}n} {et~al.}(2002){Beltr{\'a}n}, {Girart}, {Estalella}, {Ho}, \& {Palau}}]{Beltran2002}
{Beltr{\'a}n}, M.~T., {Girart}, J.~M., {Estalella}, R., {Ho}, P. T.~P., \& {Palau}, A. 2002, \apj, 573, 246

\bibitem[{{Blandford} \& {Payne}(1982)}]{Blandford1982}
{Blandford}, R.~D. \& {Payne}, D.~G. 1982, \mnras, 199, 883

\bibitem[{{Busquet} {et~al.}(2019){Busquet}, {Girart}, {Estalella}, {Fern{\'a}ndez-L{\'o}pez}, {Galv{\'a}n-Madrid}, {Anglada}, {Carrasco-Gonz{\'a}lez}, {A{\~n}ez-L{\'o}pez}, {Curiel}, {Osorio}, {Rodr{\'\i}guez}, \& {Torrelles}}]{Busquet2019}
{Busquet}, G., {Girart}, J.~M., {Estalella}, R., {et~al.} 2019, \aap, 623, L8

\bibitem[{{Carrasco-Gonz{\'a}lez} {et~al.}(2012){Carrasco-Gonz{\'a}lez}, {Galv{\'a}n-Madrid}, {Anglada}, {Osorio}, {D'Alessio}, {Hofner}, {Rodr{\'\i}guez}, {Linz}, \& {Araya}}]{Carrasco2012}
{Carrasco-Gonz{\'a}lez}, C., {Galv{\'a}n-Madrid}, R., {Anglada}, G., {et~al.} 2012, \apjl, 752, L29

\bibitem[{{CASA Team} {et~al.}(2022){CASA Team}, {Bean}, {Bhatnagar}, {Castro}, {Donovan Meyer}, {Emonts}, {Garcia}, {Garwood}, {Golap}, {Gonzalez Villalba}, {Harris}, {Hayashi}, {Hoskins}, {Hsieh}, {Jagannathan}, {Kawasaki}, {Keimpema}, {Kettenis}, {Lopez}, {Marvil}, {Masters}, {McNichols}, {Mehringer}, {Miel}, {Moellenbrock}, {Montesino}, {Nakazato}, {Ott}, {Petry}, {Pokorny}, {Raba}, {Rau}, {Schiebel}, {Schweighart}, {Sekhar}, {Shimada}, {Small}, {Steeb}, {Sugimoto}, {Suoranta}, {Tsutsumi}, {van Bemmel}, {Verkouter}, {Wells}, {Xiong}, {Szomoru}, {Griffith}, {Glendenning}, \& {Kern}}]{CASAteam}
{CASA Team}, {Bean}, B., {Bhatnagar}, S., {et~al.} 2022, \pasp, 134, 114501

\bibitem[{{Caselli} {et~al.}(1995){Caselli}, {Myers}, \& {Thaddeus}}]{Caselli1995}
{Caselli}, P., {Myers}, P.~C., \& {Thaddeus}, P. 1995, \apjl, 455, L77

\bibitem[{{Cunningham} {et~al.}(2009){Cunningham}, {Moeckel}, \& {Bally}}]{Cunningham2009}
{Cunningham}, N.~J., {Moeckel}, N., \& {Bally}, J. 2009, \apj, 692, 943

\bibitem[{{de Valon} {et~al.}(2020){de Valon}, {Dougados}, {Cabrit}, {Louvet}, {Zapata}, \& {Mardones}}]{dValon2020}
{de Valon}, A., {Dougados}, C., {Cabrit}, S., {et~al.} 2020, \aap, 634, L12

\bibitem[{{de Valon} {et~al.}(2022){de Valon}, {Dougados}, {Cabrit}, {Louvet}, {Zapata}, \& {Mardones}}]{dValon2022}
{de Valon}, A., {Dougados}, C., {Cabrit}, S., {et~al.} 2022, \aap, 668, A78

\bibitem[{{Estalella}(2017)}]{2017Estalella}
{Estalella}, R. 2017, \pasp, 129, 025003

\bibitem[{{Fern{\'a}ndez-L{\'o}pez} {et~al.}(2011{\natexlab{a}}){Fern{\'a}ndez-L{\'o}pez}, {Curiel}, {Girart}, {Ho}, {Patel}, \& {G{\'o}mez}}]{FernandezLopez2011a}
{Fern{\'a}ndez-L{\'o}pez}, M., {Curiel}, S., {Girart}, J.~M., {et~al.} 2011{\natexlab{a}}, \aj, 141, 72

\bibitem[{{Fern{\'a}ndez-L{\'o}pez} {et~al.}(2011{\natexlab{b}}){Fern{\'a}ndez-L{\'o}pez}, {Girart}, {Curiel}, {G{\'o}mez}, {Ho}, \& {Patel}}]{FernandezLopez2011b}
{Fern{\'a}ndez-L{\'o}pez}, M., {Girart}, J.~M., {Curiel}, S., {et~al.} 2011{\natexlab{b}}, \aj, 142, 97

\bibitem[{{Fern{\'a}ndez-L{\'o}pez} {et~al.}(2013){Fern{\'a}ndez-L{\'o}pez}, {Girart}, {Curiel}, {Zapata}, {Fonfr{\'\i}a}, \& {Qiu}}]{FernandezLopez2013}
{Fern{\'a}ndez-L{\'o}pez}, M., {Girart}, J.~M., {Curiel}, S., {et~al.} 2013, \apj, 778, 72

\bibitem[{{Fern{\'a}ndez-L{\'o}pez} {et~al.}(2023){Fern{\'a}ndez-L{\'o}pez}, {Girart}, {L{\'o}pez-V{\'a}zquez}, {Estalella}, {Busquet}, {Curiel}, \& {A{\~n}ez-L{\'o}pez}}]{FernandezLopez2023}
{Fern{\'a}ndez-L{\'o}pez}, M., {Girart}, J.~M., {L{\'o}pez-V{\'a}zquez}, J.~A., {et~al.} 2023, \apj, 956, 82

\bibitem[{{Fern{\'a}ndez-L{\'o}pez} {et~al.}(2020){Fern{\'a}ndez-L{\'o}pez}, {Zapata}, {Rodr{\'\i}guez}, {Vazzano}, {Guzm{\'a}n}, \& {L{\'o}pez}}]{FernandezLopez2020}
{Fern{\'a}ndez-L{\'o}pez}, M., {Zapata}, L.~A., {Rodr{\'\i}guez}, L.~F., {et~al.} 2020, \aj, 159, 171

\bibitem[{{Gerner} {et~al.}(2014){Gerner}, {Beuther}, {Semenov}, {Linz}, {Vasyunina}, {Bihr}, {Shirley}, \& {Henning}}]{Gerner2014}
{Gerner}, T., {Beuther}, H., {Semenov}, D., {et~al.} 2014, \aap, 563, A97

\bibitem[{{Gieser} {et~al.}(2021){Gieser}, {Beuther}, {Semenov}, {Ahmadi}, {Suri}, {M{\"o}ller}, {Beltr{\'a}n}, {Klaassen}, {Zhang}, {Urquhart}, {Henning}, {Feng}, {Galv{\'a}n-Madrid}, {de Souza Magalh{\~a}es}, {Moscadelli}, {Longmore}, {Leurini}, {Kuiper}, {Peters}, {Menten}, {Csengeri}, {Fuller}, {Wyrowski}, {Lumsden}, {S{\'a}nchez-Monge}, {Maud}, {Linz}, {Palau}, {Schilke}, {Pety}, {Pudritz}, {Winters}, \& {Pi{\'e}tu}}]{Gieser2021}
{Gieser}, C., {Beuther}, H., {Semenov}, D., {et~al.} 2021, \aap, 648, A66

\bibitem[{{Girart} {et~al.}(2018){Girart}, {Fern{\'a}ndez-L{\'o}pez}, {Li}, {Yang}, {Estalella}, {Anglada}, {{\'A}{\~n}ez-L{\'o}pez}, {Busquet}, {Carrasco-Gonz{\'a}lez}, {Curiel}, {Galvan-Madrid}, {G{\'o}mez}, {de Gregorio-Monsalvo}, {Jim{\'e}nez-Serra}, {Krasnopolsky}, {Mart{\'\i}}, {Osorio}, {Padovani}, {Rao}, {Rodr{\'\i}guez}, \& {Torrelles}}]{Girart2018}
{Girart}, J.~M., {Fern{\'a}ndez-L{\'o}pez}, M., {Li}, Z.~Y., {et~al.} 2018, \apjl, 856, L27

\bibitem[{{G{\'o}mez} {et~al.}(2003){G{\'o}mez}, {Rodr{\'\i}guez}, {Girart}, {Garay}, \& {Mart{\'\i}}}]{Gomez2003}
{G{\'o}mez}, Y., {Rodr{\'\i}guez}, L.~F., {Girart}, J.~M., {Garay}, G., \& {Mart{\'\i}}, J. 2003, \apj, 597, 414

\bibitem[{{Hartigan}(2003)}]{Hartigan2003}
{Hartigan}, P. 2003, \apss, 287, 111

\bibitem[{{Hartigan} {et~al.}(2000){Hartigan}, {Bally}, {Reipurth}, \& {Morse}}]{Hartigan2000}
{Hartigan}, P., {Bally}, J., {Reipurth}, B., \& {Morse}, J.~A. 2000, in Protostars and Planets IV, ed. V.~{Mannings}, A.~P. {Boss}, \& S.~S. {Russell}, 841

\bibitem[{{Hartigan} {et~al.}(1987){Hartigan}, {Raymond}, \& {Hartmann}}]{Hartigan1987}
{Hartigan}, P., {Raymond}, J., \& {Hartmann}, L. 1987, \apj, 316, 323

\bibitem[{{Heathcote} {et~al.}(1998){Heathcote}, {Reipurth}, \& {Raga}}]{Heathcote1998}
{Heathcote}, S., {Reipurth}, B., \& {Raga}, A.~C. 1998, \aj, 116, 1940

\bibitem[{{J{\o}rgensen} {et~al.}(2004){J{\o}rgensen}, {Hogerheijde}, {van Dishoeck}, {Blake}, \& {Sch{\"o}ier}}]{Jorgensen2004}
{J{\o}rgensen}, J.~K., {Hogerheijde}, M.~R., {van Dishoeck}, E.~F., {Blake}, G.~A., \& {Sch{\"o}ier}, F.~L. 2004, \aap, 413, 993

\bibitem[{{Kajdi{\v{c}}} {et~al.}(2012){Kajdi{\v{c}}}, {Reipurth}, {Raga}, {Bally}, \& {Walawender}}]{Kajdic2012}
{Kajdi{\v{c}}}, P., {Reipurth}, B., {Raga}, A.~C., {Bally}, J., \& {Walawender}, J. 2012, \aj, 143, 106

\bibitem[{{Kurtz} \& {Hofner}(2005)}]{Kurtz2005}
{Kurtz}, S. \& {Hofner}, P. 2005, \aj, 130, 711

\bibitem[{{Kurtz} {et~al.}(2004){Kurtz}, {Hofner}, \& {{\'A}lvarez}}]{Kurtz2004}
{Kurtz}, S., {Hofner}, P., \& {{\'A}lvarez}, C.~V. 2004, \apjs, 155, 149

\bibitem[{{Kwon} {et~al.}(2015){Kwon}, {Fern{\'a}ndez-L{\'o}pez}, {Stephens}, \& {Looney}}]{Kwon2015}
{Kwon}, W., {Fern{\'a}ndez-L{\'o}pez}, M., {Stephens}, I.~W., \& {Looney}, L.~W. 2015, \apj, 814, 43

\bibitem[{{Lee} {et~al.}(2000){Lee}, {Mundy}, {Reipurth}, {Ostriker}, \& {Stone}}]{Lee2000}
{Lee}, C.-F., {Mundy}, L.~G., {Reipurth}, B., {Ostriker}, E.~C., \& {Stone}, J.~M. 2000, \apj, 542, 925

\bibitem[{{Lee} {et~al.}(2001){Lee}, {Stone}, {Ostriker}, \& {Mundy}}]{Lee2001}
{Lee}, C.-F., {Stone}, J.~M., {Ostriker}, E.~C., \& {Mundy}, L.~G. 2001, \apj, 557, 429

\bibitem[{{L{\'o}pez} {et~al.}(2022){L{\'o}pez}, {Estalella}, {Beltr{\'a}n}, {Massi}, {Acosta-Pulido}, \& {Girart}}]{2022Lopez}
{L{\'o}pez}, R., {Estalella}, R., {Beltr{\'a}n}, M.~T., {et~al.} 2022, \aap, 661, A106

\bibitem[{{L{\'o}pez-Sepulcre} {et~al.}(2009){L{\'o}pez-Sepulcre}, {Codella}, {Cesaroni}, {Marcelino}, \& {Walmsley}}]{LopezSepulcre2009}
{L{\'o}pez-Sepulcre}, A., {Codella}, C., {Cesaroni}, R., {Marcelino}, N., \& {Walmsley}, C.~M. 2009, \aap, 499, 811

\bibitem[{{L{\'o}pez-V{\'a}zquez} {et~al.}(2024){L{\'o}pez-V{\'a}zquez}, {Lee}, {Fern{\'a}ndez-L{\'o}pez}, {Louvet}, {Guerra-Alvarado}, \& {Zapata}}]{JALV2024}
{L{\'o}pez-V{\'a}zquez}, J.~A., {Lee}, C.-F., {Fern{\'a}ndez-L{\'o}pez}, M., {et~al.} 2024, \apj, 962, 28

\bibitem[{{Mangum} \& {Shirley}(2015)}]{Mangum2015}
{Mangum}, J.~G. \& {Shirley}, Y.~L. 2015, \pasp, 127, 266

\bibitem[{{Marti} {et~al.}(1993){Marti}, {Rodriguez}, \& {Reipurth}}]{Marti1993}
{Marti}, J., {Rodriguez}, L.~F., \& {Reipurth}, B. 1993, \apj, 416, 208

\bibitem[{{Mart{\'\i}} {et~al.}(1999){Mart{\'\i}}, {Rodr{\'\i}guez}, \& {Torrelles}}]{Marti1999}
{Mart{\'\i}}, J., {Rodr{\'\i}guez}, L.~F., \& {Torrelles}, J.~M. 1999, \aap, 345, L5

\bibitem[{{Masqu{\'e}} {et~al.}(2012){Masqu{\'e}}, {Girart}, {Estalella}, {Rodr{\'\i}guez}, \& {Beltr{\'a}n}}]{Masque2012}
{Masqu{\'e}}, J.~M., {Girart}, J.~M., {Estalella}, R., {Rodr{\'\i}guez}, L.~F., \& {Beltr{\'a}n}, M.~T. 2012, \apjl, 758, L10

\bibitem[{{Masqu{\'e}} {et~al.}(2015){Masqu{\'e}}, {Rodr{\'\i}guez}, {Araudo}, {Estalella}, {Carrasco-Gonz{\'a}lez}, {Anglada}, {Girart}, \& {Osorio}}]{Masque2015}
{Masqu{\'e}}, J.~M., {Rodr{\'\i}guez}, L.~F., {Araudo}, A., {et~al.} 2015, \apj, 814, 44

\bibitem[{{Maud} {et~al.}(2015){Maud}, {Moore}, {Lumsden}, {Mottram}, {Urquhart}, \& {Hoare}}]{Maud2015}
{Maud}, L.~T., {Moore}, T.~J.~T., {Lumsden}, S.~L., {et~al.} 2015, \mnras, 453, 645

\bibitem[{{Maureira} {et~al.}(2017){Maureira}, {Arce}, {Dunham}, {Pineda}, {Fern{\'a}ndez-L{\'o}pez}, {Chen}, \& {Mardones}}]{Maureira2017}
{Maureira}, M.~J., {Arce}, H.~G., {Dunham}, M.~M., {et~al.} 2017, \apj, 838, 60

\bibitem[{{M{\"o}ller} {et~al.}(2017){M{\"o}ller}, {Endres}, \& {Schilke}}]{Moller2017}
{M{\"o}ller}, T., {Endres}, C., \& {Schilke}, P. 2017, \aap, 598, A7

\bibitem[{{Moscadelli} {et~al.}(2023){Moscadelli}, {Oliva}, {Surcis}, {Sanna}, {Beltr{\'a}n}, \& {Kuiper}}]{Moscadelli2023}
{Moscadelli}, L., {Oliva}, A., {Surcis}, G., {et~al.} 2023, \aap, 680, A107

\bibitem[{{Moscadelli} {et~al.}(2022){Moscadelli}, {Sanna}, {Beuther}, {Oliva}, \& {Kuiper}}]{Moscadelli2022}
{Moscadelli}, L., {Sanna}, A., {Beuther}, H., {Oliva}, A., \& {Kuiper}, R. 2022, Nature Astronomy, 6, 1068

\bibitem[{{M{\"u}ller} {et~al.}(2005){M{\"u}ller}, {Schl{\"o}der}, {Stutzki}, \& {Winnewisser}}]{2005Muller}
{M{\"u}ller}, H. S.~P., {Schl{\"o}der}, F., {Stutzki}, J., \& {Winnewisser}, G. 2005, Journal of Molecular Structure, 742, 215

\bibitem[{{Nakamura} \& {Li}(2014)}]{Nakamura2014}
{Nakamura}, F. \& {Li}, Z.-Y. 2014, \apj, 783, 115

\bibitem[{{Newville} {et~al.}(2020){Newville}, {Otten}, {Nelson}, {Ingargiola}, {Stensitzki}, {Allan}, {Fox}, {Carter}, {Micha{\l}}, {Pustakhod}, {Ram}, {Glenn}, {Deil}, {Stuermer}, {Beelen}, {Frost}, {Zobrist}, {Mark}, {Pasquevich}, {Hansen}, {Spillane}, {Caldwell}, {Polloreno}, {Andrewhannum}, {Fraine}, {Deep-42-Thought}, {Maier}, {Gamari}, {Persaud}, \& {Almarza}}]{Newville2020}
{Newville}, M., {Otten}, R., {Nelson}, A., {et~al.} 2020, {lmfit/lmfit-py 1.0.1, Zenodo, doi:10.5281/zenodo.598352}

\bibitem[{{Plunkett} {et~al.}(2015){Plunkett}, {Arce}, {Corder}, {Dunham}, {Garay}, \& {Mardones}}]{Plunkett2015}
{Plunkett}, A.~L., {Arce}, H.~G., {Corder}, S.~A., {et~al.} 2015, \apj, 803, 22

\bibitem[{{Podio} {et~al.}(2016){Podio}, {Codella}, {Gueth}, {Cabrit}, {Maury}, {Tabone}, {Lef{\`e}vre}, {Anderl}, {Andr{\'e}}, {Belloche}, {Bontemps}, {Hennebelle}, {Lefloch}, {Maret}, \& {Testi}}]{Podio2016}
{Podio}, L., {Codella}, C., {Gueth}, F., {et~al.} 2016, \aap, 593, L4

\bibitem[{{Podio} {et~al.}(2014){Podio}, {Lefloch}, {Ceccarelli}, {Codella}, \& {Bachiller}}]{Podio2014}
{Podio}, L., {Lefloch}, B., {Ceccarelli}, C., {Codella}, C., \& {Bachiller}, R. 2014, \aap, 565, A64

\bibitem[{{Pudritz} \& {Norman}(1983)}]{Pudritz1983}
{Pudritz}, R.~E. \& {Norman}, C.~A. 1983, \apj, 274, 677

\bibitem[{{Purcell} {et~al.}(2006){Purcell}, {Balasubramanyam}, {Burton}, {Walsh}, {Minier}, {Hunt-Cunningham}, {Kedziora-Chudczer}, {Longmore}, {Hill}, {Bains}, {Barnes}, {Busfield}, {Calisse}, {Crighton}, {Curran}, {Davis}, {Dempsey}, {Derragopian}, {Fulton}, {Hidas}, {Hoare}, {Lee}, {Ladd}, {Lumsden}, {Moore}, {Murphy}, {Oudmaijer}, {Pracy}, {Rathborne}, {Robertson}, {Schultz}, {Shobbrook}, {Sparks}, {Storey}, \& {Travouillion}}]{Purcell2006}
{Purcell}, C.~R., {Balasubramanyam}, R., {Burton}, M.~G., {et~al.} 2006, \mnras, 367, 553

\bibitem[{{Qiu} \& {Zhang}(2009)}]{Qiu2009}
{Qiu}, K. \& {Zhang}, Q. 2009, \apjl, 702, L66

\bibitem[{{Raga} {et~al.}(1998){Raga}, {Canto}, {Curiel}, \& {Taylor}}]{Raga1998}
{Raga}, A.~C., {Canto}, J., {Curiel}, S., \& {Taylor}, S. 1998, \mnras, 295, 738

\bibitem[{{Raga} {et~al.}(2002){Raga}, {de Gouveia Dal Pino}, {Noriega-Crespo}, {Mininni}, \& {Vel{\'a}zquez}}]{Raga2002}
{Raga}, A.~C., {de Gouveia Dal Pino}, E.~M., {Noriega-Crespo}, A., {Mininni}, P.~D., \& {Vel{\'a}zquez}, P.~F. 2002, \aap, 392, 267

\bibitem[{{Raga} \& {Wang}(1994)}]{Raga1994}
{Raga}, A.~C. \& {Wang}, L. 1994, \mnras, 268, 354

\bibitem[{{Remijan} {et~al.}(2007){Remijan}, {Markwick-Kemper}, \& {ALMA Working Group on Spectral Line Frequencies}}]{Remijan2007}
{Remijan}, A.~J., {Markwick-Kemper}, A., \& {ALMA Working Group on Spectral Line Frequencies}. 2007, in American Astronomical Society Meeting Abstracts, Vol. 211, American Astronomical Society Meeting Abstracts, 132.11

\bibitem[{{Rodr{\'\i}guez} \& {Reipurth}(1989)}]{Rodriguez1989}
{Rodr{\'\i}guez}, L.~F. \& {Reipurth}, B. 1989, \rmxaa, 17, 59

\bibitem[{{Sai} {et~al.}(2024){Sai}, {Yen}, {Machida}, {Ohashi}, {Aso}, {Maury}, \& {Maret}}]{Sai2024}
{Sai}, J., {Yen}, H.-W., {Machida}, M.~N., {et~al.} 2024, arXiv e-prints, arXiv:2403.10857

\bibitem[{{S{\'a}nchez-Monge} {et~al.}(2013){S{\'a}nchez-Monge}, {L{\'o}pez-Sepulcre}, {Cesaroni}, {Walmsley}, {Codella}, {Beltr{\'a}n}, {Pestalozzi}, \& {Molinari}}]{SanchezMonge2013}
{S{\'a}nchez-Monge}, {\'A}., {L{\'o}pez-Sepulcre}, A., {Cesaroni}, R., {et~al.} 2013, \aap, 557, A94

\bibitem[{{Shirley}(2015)}]{Shirley2015}
{Shirley}, Y.~L. 2015, \pasp, 127, 299

\bibitem[{{Shirley} {et~al.}(2005){Shirley}, {Nordhaus}, {Grcevich}, {Evans}, {Rawlings}, \& {Tatematsu}}]{Shirley2005}
{Shirley}, Y.~L., {Nordhaus}, M.~K., {Grcevich}, J.~M., {et~al.} 2005, \apj, 632, 982

\bibitem[{{Shu} {et~al.}(1991){Shu}, {Ruden}, {Lada}, \& {Lizano}}]{Shu1991}
{Shu}, F.~H., {Ruden}, S.~P., {Lada}, C.~J., \& {Lizano}, S. 1991, \apjl, 370, L31

\bibitem[{{Snell} {et~al.}(1980){Snell}, {Loren}, \& {Plambeck}}]{Snell1980}
{Snell}, R.~L., {Loren}, R.~B., \& {Plambeck}, R.~L. 1980, \apjl, 239, L17

\bibitem[{{Soker} \& {Mcley}(2013)}]{Soker2013}
{Soker}, N. \& {Mcley}, L. 2013, \apjl, 772, L22

\bibitem[{{Stanke} {et~al.}(2022){Stanke}, {Arce}, {Bally}, {Bergman}, {Carpenter}, {Davis}, {Dent}, {Di Francesco}, {Eisl{\"o}ffel}, {Froebrich}, {Ginsburg}, {Heyer}, {Johnstone}, {Mardones}, {McCaughrean}, {Megeath}, {Nakamura}, {Smith}, {Stutz}, {Tatematsu}, {Walker}, {Williams}, {Zinnecker}, {Swift}, {Kulesa}, {Peters}, {Duffy}, {Kloosterman}, {Y{\ensuremath{\i}}ld{\ensuremath{\i}}z}, {Pineda}, {De Breuck}, \& {Klein}}]{Stanke2022}
{Stanke}, T., {Arce}, H.~G., {Bally}, J., {et~al.} 2022, \aap, 658, A178

\bibitem[{{Takaishi} {et~al.}(2024){Takaishi}, {Tsukamoto}, {Kido}, {Takakuwa}, {Misugi}, {Kudoh}, \& {Suto}}]{Takaishi2024}
{Takaishi}, D., {Tsukamoto}, Y., {Kido}, M., {et~al.} 2024, arXiv e-prints, arXiv:2401.02525

\bibitem[{{Tang} {et~al.}(2018){Tang}, {Henkel}, {Wyrowski}, {Giannetti}, {Menten}, {Csengeri}, {Leurini}, {Urquhart}, {K{\"o}nig}, {G{\"u}sten}, {Lin}, {Zheng}, {Esimbek}, \& {Zhou}}]{Tang2018}
{Tang}, X.~D., {Henkel}, C., {Wyrowski}, F., {et~al.} 2018, \aap, 611, A6

\bibitem[{{Teixeira} {et~al.}(2008){Teixeira}, {McCoey}, {Fich}, \& {Lada}}]{Teixeira2008}
{Teixeira}, P.~S., {McCoey}, C., {Fich}, M., \& {Lada}, C.~J. 2008, \mnras, 384, 71

\bibitem[{{Tobin} {et~al.}(2012){Tobin}, {Hartmann}, {Bergin}, {Chiang}, {Looney}, {Chandler}, {Maret}, \& {Heitsch}}]{Tobin2012}
{Tobin}, J.~J., {Hartmann}, L., {Bergin}, E., {et~al.} 2012, \apj, 748, 16

\bibitem[{{Toledano-Ju{\'a}rez} {et~al.}(2023){Toledano-Ju{\'a}rez}, {de la Fuente}, {Trinidad}, {Tafoya}, \& {Nigoche-Netro}}]{Toledano2023}
{Toledano-Ju{\'a}rez}, I., {de la Fuente}, E., {Trinidad}, M.~A., {Tafoya}, D., \& {Nigoche-Netro}, A. 2023, \mnras, 522, 1591

\bibitem[{{Vazzano} {et~al.}(2021){Vazzano}, {Fern{\'a}ndez-L{\'o}pez}, {Plunkett}, {de Gregorio-Monsalvo}, {Santamar{\'\i}a-Miranda}, {Takahashi}, \& {Lopez}}]{Vazzano2021}
{Vazzano}, M.~M., {Fern{\'a}ndez-L{\'o}pez}, M., {Plunkett}, A., {et~al.} 2021, \aap, 648, A41

\bibitem[{{Zapata} {et~al.}(2013){Zapata}, {Fernandez-Lopez}, {Curiel}, {Patel}, \& {Rodriguez}}]{Zapata2013}
{Zapata}, L.~A., {Fernandez-Lopez}, M., {Curiel}, S., {Patel}, N., \& {Rodriguez}, L.~F. 2013, arXiv e-prints, arXiv:1305.4084

\bibitem[{{Zapata} {et~al.}(2018){Zapata}, {Fern{\'a}ndez-L{\'o}pez}, {Rodr{\'\i}guez}, {Garay}, {Takahashi}, {Lee}, \& {Hern{\'a}ndez-G{\'o}mez}}]{Zapata2018}
{Zapata}, L.~A., {Fern{\'a}ndez-L{\'o}pez}, M., {Rodr{\'\i}guez}, L.~F., {et~al.} 2018, \aj, 156, 239

\bibitem[{{Zhang} {et~al.}(2019){Zhang}, {Arce}, {Mardones}, {Cabrit}, {Dunham}, {Garay}, {Noriega-Crespo}, {Offner}, {Raga}, \& {Corder}}]{Zhang2019}
{Zhang}, Y., {Arce}, H.~G., {Mardones}, D., {et~al.} 2019, \apj, 883, 1

\bibitem[{{Zhao} {et~al.}(2024){Zhao}, {Tang}, {Henkel}, {Gong}, {Lin}, {Li}, {He}, {Ao}, {Lu}, {Liu}, {Sun}, {Wang}, {Chen}, {Esimbek}, {Zhou}, {Wu}, {Qiu}, {Zheng}, {Li}, {Luo}, \& {Zhao}}]{Zhao2024}
{Zhao}, X., {Tang}, X.~D., {Henkel}, C., {et~al.} 2024, \aap, 687, A207

\end{thebibliography}


\begin{appendix}
\section{Channel maps}
\label{app:channel}

\begin{figure*}[t!]
\centering
\includegraphics[scale=0.6]{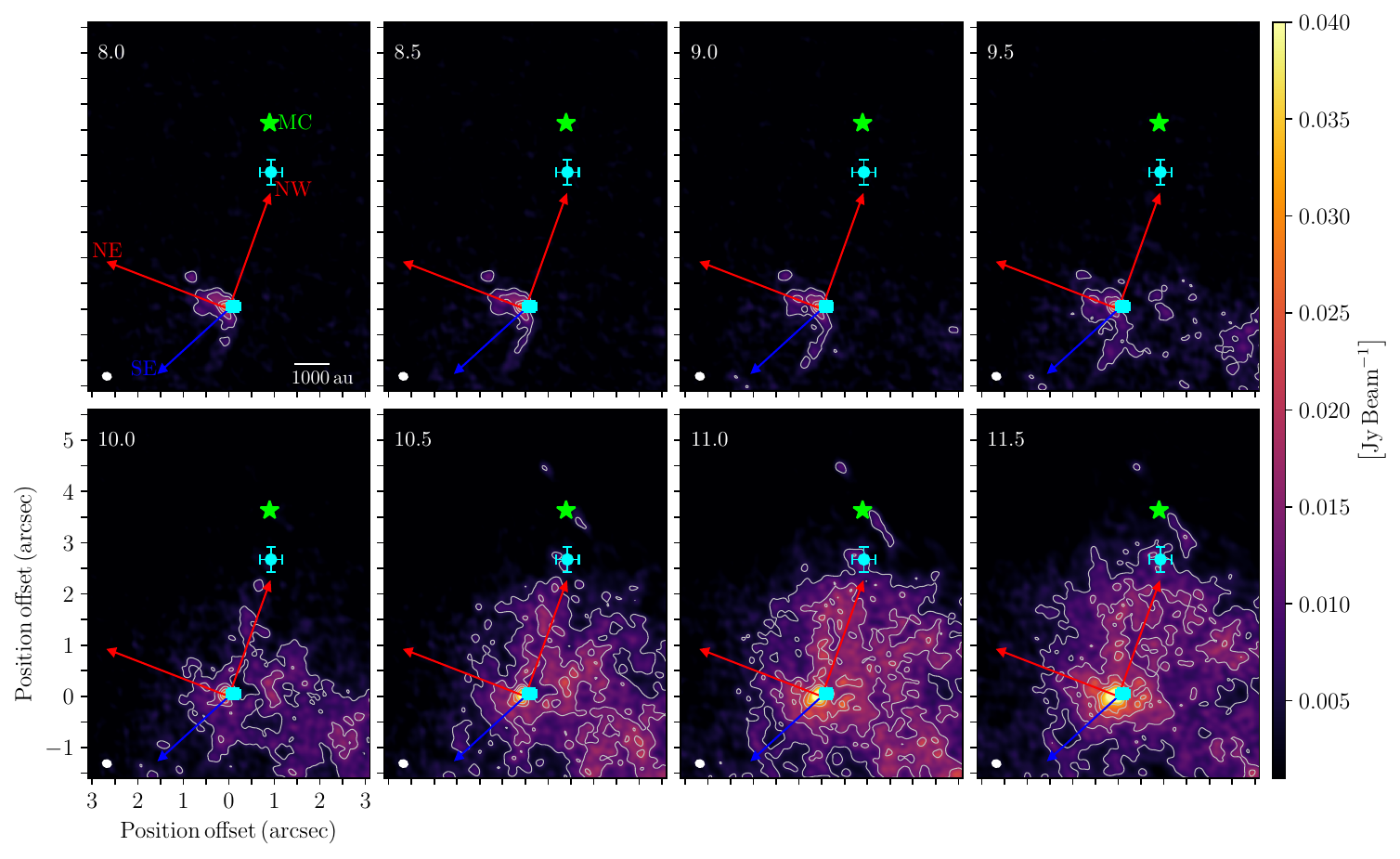}
\caption{Blueshifted channel maps of the H$_2$CO molecular line emission of the GGD 27--MM2(E) protostellar system. The channel velocity in \kms is indicated in the left top corner. The contours start at 3$\sigma$ in steps of 3$\sigma$, 6$\sigma$, 9$\sigma$, and 12$\sigma$, where $\sigma=2.1$ mJy beam$^{-1}$. Cyan dot and square mark the position of the CH$_3$OH and H$_2$O masers sources obtained from \citet{Kurtz2004}, and \citet{Kurtz2005}, respectively. Green star presents the position of the source MC taken from \citet{Qiu2009}. The synthesized beam in all panels is shown in the left lower corner. The two red arrows present the redshifted NE and NW CO outflow directions, while the blue arrow is the blueshifted SE CO outflow direction taken from \citet{FernandezLopez2013}.}
\label{fig:bluechannels}
\end{figure*}

\begin{figure*}[t!]
\centering
\includegraphics[scale=0.6]{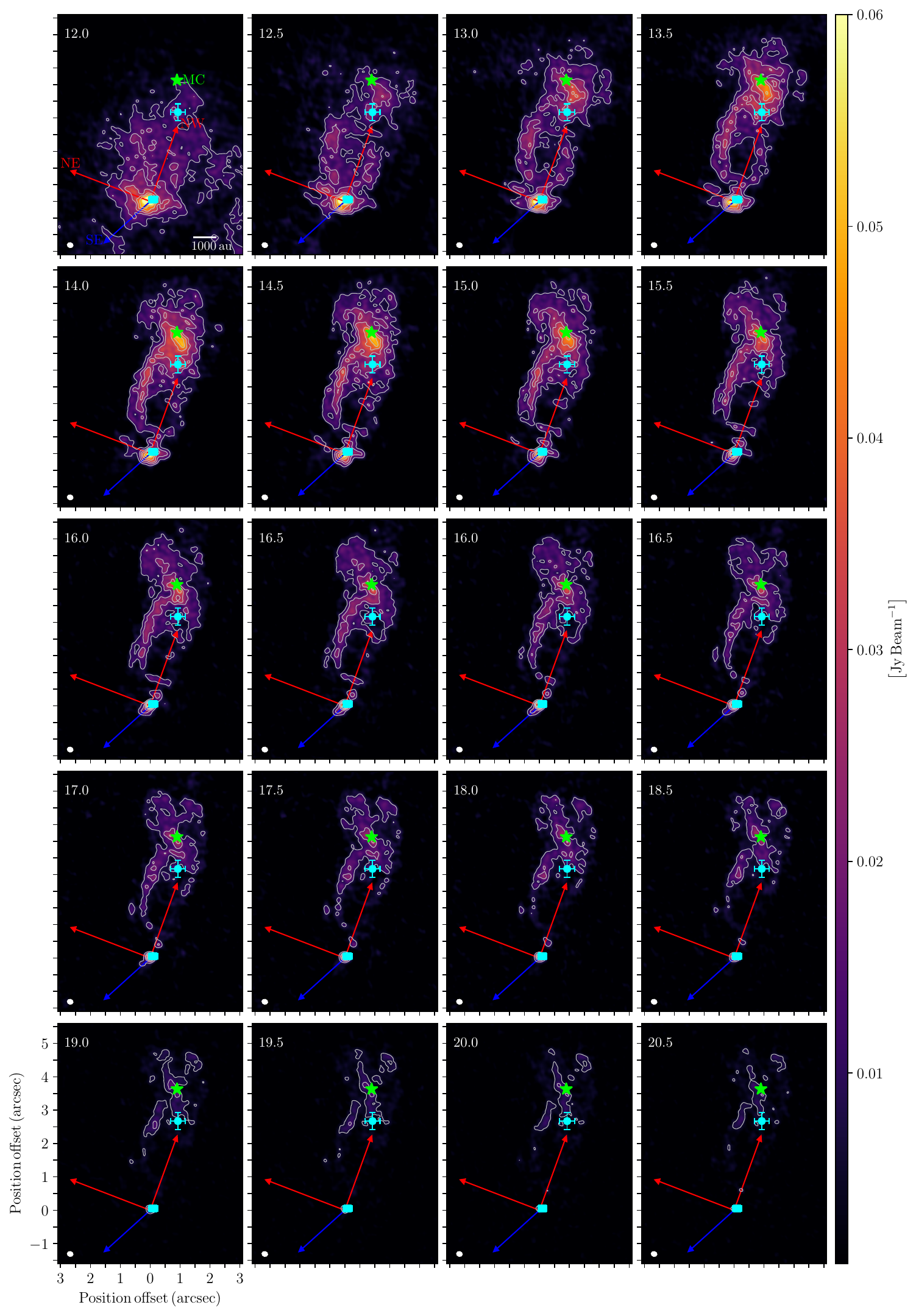}
\caption{Redshifted channel maps of the H$_2$CO molecular line emission of the GGD 27--MM2(E) protostellar system. Same description as \ref{fig:bluechannels}.}
\label{fig:redchannels}
\end{figure*}

In this appendix Figures \ref{fig:bluechannels} and \ref{fig:redchannels} show the complete velocity cube of the emission of the H$_2$CO molecular line with a channel width of 0.5 \kms. The arrows in both Figures represent the direction of the CO molecular outflow, the redshifted northeast and northwest (red arrows) and blueshifted southeast (blue arrow).

\end{appendix}

\end{document}